\documentclass[12pt, draftclsnofoot, onecolumn]{IEEEtran}

\usepackage{caption}
\usepackage{subfigure}
\usepackage{graphicx,amssymb,amsmath}
\usepackage{multicol}
\usepackage[noadjust]{cite}
\usepackage{setspace}
\usepackage{stfloats}
\usepackage{amsthm}
\usepackage{flushend}
\usepackage{cases,subeqnarray}
\usepackage{bm,multirow,bigstrut}
\usepackage{textcomp}
\usepackage{latexsym,bm}
\usepackage{bm}
\usepackage{booktabs,changebar}
\usepackage{xcolor}
\usepackage{mathtools}
\usepackage{dsfont}
\usepackage{extarrows}
\usepackage{hyperref}
\usepackage{colortbl}
\definecolor{tabcolor}{rgb}{.105,.410,.113}
\makeatletter

\newcommand{\Rmnum}[1]{\expandafter\@slowromancap\romannumeral #1@}
\makeatother

\theoremstyle{plain}

\theoremstyle{remark}
\newtheorem{remark}{Remark}
\newtheorem{coro}{Corollary}

\IEEEoverridecommandlockouts

\makeatother
\begin{document}
\title{Reconfigurable Intelligent Surfaces with Outdated Channel State Information: Centralized vs. Distributed Deployments}

\author{Yan Zhang, Jiayi~Zhang,~\IEEEmembership{Senior Member,~IEEE}, Marco~Di~Renzo,~\IEEEmembership{Fellow,~IEEE}
\\Huahua Xiao, and Bo Ai,~\IEEEmembership{Fellow,~IEEE}
\thanks{Y.~Zhang and J. Zhang are with the School of Electronic and Information Engineering, Beijing Jiaotong University, Beijing 100044, China.}
\thanks{M. Di Renzo is with Universit\'e Paris-Saclay, CNRS, CentraleSup\'elec, Laboratoire des Signaux et Syst\`emes, 3 Rue Joliot-Curie, 91192 Gif-sur-Yvette, France.}
\thanks{H. Xiao is with ZTE Corporation, and State Key Laboratory of Mobile Network and Mobile Multimedia Technology, Shenzhen 518057, China.}
\thanks{B. Ai is with the State Key Laboratory of Rail Traffic Control and Safety, Beijing Jiaotong University, Beijing 100044, China.}
}

\maketitle
\vspace{-1.5cm}
\begin{abstract}
In this paper, we investigate the performance of an RIS-aided wireless communication system subject to
outdated channel state information that may operate in both the near- and far-field regions.
In particular, we take two RIS deployment strategies into consideration: (i) the centralized deployment,
where all the reflecting elements are installed on a single RIS and (ii) the distributed deployment,
where the same number of reflecting elements are placed on multiple RISs.
For both deployment strategies, we derive accurate closed-form approximations for the ergodic capacity,
and we introduce tight upper and lower bounds for the ergodic capacity to obtain useful design insights.
From this analysis, we unveil that an increase of the transmit power, the Rician-$K$ factor,
the accuracy of the channel state information and the number of reflecting elements help improve the system performance.
Moreover, we prove that the centralized RIS-aided deployment may achieve a higher ergodic capacity as compared with the
distributed RIS-aided deployment when the RIS is located near the base station or near the user.
In different setups, on the other hand, we prove that the distributed deployment outperforms the centralized deployment.
Finally, the analytical results are verified by using Monte Carlo simulations.
\end{abstract}

\IEEEpeerreviewmaketitle
\section{Introduction}
A reconfigurable intelligent surface (RIS) is an artificial planar structure with integrated electronic circuits,
which is equipped with a large number of passive and low-cost scattering elements that can effectively control the
wireless propagation environment \cite{zhang2020prospective}.
By intelligently adapting the phase shifts and the amplitude response of the scattering elements of an RIS,
the signals reflected from it can be added constructively or destructively with other signals so as to enhance
the signal strength or to suppress the co-channel interference at the receiver
\cite{QQWuBeam,renzo2020smart,basar2019wireless,huang2019holographic,Liu2020ReconfigurableIS,chen2020resource,
yang2020performance,renzo2019smartAI,ntontin2019reconfigurable}.
Thanks to these properties, RISs are considered to be a promising candidate technology for future wireless communication systems.

Several works have investigated the performance of single RIS-aided wireless systems
\cite{LiangYang2020, van2020coverage, Yang2020AccurateCA, salhab2021accurate,kudathanthirige2020performance,
lin2020reconfigurable,
wang2020chernoff, wang2020study,
FerreiaBER, trigui2021bit, Badiu2020CommunicationTA,
DLiergodic, BAlexergodic,
xu2020reconfigurable,
qian2020beamforming,dharmawansa2020performance,
QintaoAcomprehensive,trigui2020performance,jung2020performance,
YangSecrecy2020,
du2021millimeter, atapattu2020reconfigurable,
tao2020performance,danufane2021path,zhang2021physical
}.
In \cite{LiangYang2020}, the authors investigated the coverage, the delay outage rate,
and the probability of the signal-to-noise-ratio (SNR) gain of an RIS-aided communication system in the presence of
Rayleigh fading channels by using the central limit theorem (CLT).
A similar system was considered in \cite{van2020coverage},
where the authors studied the coverage probability and the ergodic capacity (EC) by using the moment-matching method.
In \cite{Yang2020AccurateCA,salhab2021accurate, kudathanthirige2020performance},
the authors proposed accurate closed-form approximations for the outage probability (OP), error rate,
and average channel capacity of RIS-aided communication systems over Rayleigh and Rician fading channels, respectively.
In \cite{lin2020reconfigurable},
the authors derived the asymptotic OP and the achievable rate of an RIS-aided communication system over Rician fading channels.
In \cite{wang2020chernoff} and \cite{wang2020study},
the authors provided closed-form approximate expressions for the OP of an RIS-aided communication system.
In \cite{FerreiaBER}, approximated and upper bound expressions for the bit error rate (BER) were derived over
a Nakagami-$m$ fading channel.
The authors of \cite{trigui2021bit} and \cite{Badiu2020CommunicationTA} analyzed the BER of an RIS-aided system
by taking into account
the phase errors caused by the quantization of the phase shifts over Rayleigh and Nakagami fading channels, respectively.
In \cite{DLiergodic} and \cite{BAlexergodic}, the authors studied the EC of RIS-assisted communication systems.
Particularly, the impact of quantization phase errors was analyzed in \cite{ DLiergodic}.
In \cite{xu2020reconfigurable},
the minimum required number of phase quantization levels to achieve the full diversity order
in RIS-aided communication systems was obtained.
In \cite{qian2020beamforming},
the authors studied the performance of RIS-aided multiple-input multiple-output (MIMO) communication systems with phase noise.
A similar system was considered in \cite{dharmawansa2020performance},
in which the authors analyzed the OP and throughput of a two-tile RIS-aided wireless network over Rayleigh fading channels.
In \cite{QintaoAcomprehensive},
exact expressions of the OP and the EC for an RIS-aided system over Fox's-$H$ fading channels were provided.
By assuming the same generalized channel model,
the authors of \cite{trigui2020performance} analyzed the OP in the presence of phase noise.
In \cite{jung2020performance},
the authors studied the asymptotic data rate in an RIS-aided large antenna-array system by considering the channel hardening effects.
In \cite{YangSecrecy2020},
the secrecy OP of an RIS-aided communication system was characterized over Rayleigh fading channels.
In \cite{du2021millimeter},
the end-to-end SNR of an RIS-aided millimeter wave communication system was maximized
by optimizing the phase shifts of the RIS elements.
In \cite{atapattu2020reconfigurable},
the authors considered an RIS for assisting the communication between two users,
and the OP and spectral efficiency were studied over Rayleigh fading channels by using a Gamma approximation.
In \cite{tao2020performance},
a closed-form upper bound expression for the EC and an accurate approximation for the OP were derived for
transmission over Rician fading channels.
In \cite{danufane2021path},
the authors investigated the near- and far-field free-space path loss model for RIS-aided communication systems.

Although previous works have provided important contributions to analyze the performance of RIS-aided systems,
most of them can be applied to wireless systems in the presence of a single RIS.
More recently, the performance of distributed RIS-aided systems has been investigated in
\cite{yang2020outage, yildirim2020modeling, do2021multi, galappaththige2020performance, zhang2021intelligent,abrardo2020intelligent}.
In \cite{yang2020outage},
the authors investigated the OP and sum-rate of a dual-hop cooperative network assisted by the RIS which has the highest
instantaneous end-to-end SNR among multiple available RISs.
In \cite{yildirim2020modeling},
the authors studied multi-RIS-aided systems for application to non-line-of-sight indoor and outdoor communications.
In \cite{do2021multi},
the authors proposed two multi-RIS-aided schemes and provided different approximate methods to analyze the OP and the EC over
Nakagami-$m$ fading channels.
In \cite{galappaththige2020performance},
the authors studied the OP,
the average achievable rate and the average symbol error rate of a distributed RIS-aided communication system.
In \cite{zhang2021intelligent},
the authors compared the capacity region of an RIS-aided two-user communication system under centralized and distributed
RIS deployment strategies.
In \cite{abrardo2020intelligent},
the authors proposed an optimization algorithm to configure multiple RISs to maximize the sum-rate of multi-user
MIMO communication systems based on statistical channel information (CSI).
Although these works have made efforts to investigate distributed RIS-aided communication systems,
they have neither analyzed the performance in the near-field region of the RISs,
which may not be overlooked in some network deployments \cite{danufane2021path},
nor they have assessed the impact of outdated CSI,
which is critical to acquire in RIS-aided systems \cite{yu2020robustoutdated2,yang2020deepoutdated1,hashemi2021deep}.
It is worth mentioning that the majority of previous works analyzed the performance of RIS-aided communication systems
under the assumption of perfect CSI or statistical CSI to design the optimal phase shifts at the RIS.
However, acquiring accurate CSI is very challenging \cite{zappone2020overhead}.
Due to the associated feedback delay and the user mobility,
the channel learned via estimation may often be outdated.
Therefore, it is of great importance to take into account the impact of outdated CSI on the system performance.

Motivated by these considerations, in this paper, we analyze the performance of an RIS-aided communication system by considering
two RIS deployment strategies, i.e., the centralized and distributed case, by taking into account the impact of outdated CSI.
More specifically, we analyze and compare the system performance over Rician fading channels in both the near- and far-field regions
of the RISs.
To this end, we derive accurate closed-form approximate expressions for the EC, which can be used in the near- and far-field
regions of centralized and distributed RIS deployments, by using the moment-matching method to approximate the cumulative
distribution function (CDF) of end-to-end SNR with a Gamma distribution.
In order to gain design insights, in addition, we introduce tight lower and upper bounds for the EC.
With the aid of the introduced analytical frameworks, we characterize the impact of key parameters on the system performance
and compare centralized and distributed RIS deployments against each other.
In particular, we show that an increase of the transmit power, the Rician-$K$ factor, the accuracy of CSI and the size of
the unit cells results in an improved EC.
As a function of the number of RIS elements, we show that the EC increases and reaches a finite limit when the number of
RIS elements tends to infinity. This is attributed to the near-field propagation conditions in this asymptotic regime.
Furthermore, it is shown that different location deployments for the RISs and CSI accuracy result in different conclusions
when comparing the system performance of centralized and distributed deployments.
The main contributions of this paper can be summarized as follows.
\begin{itemize}
\item
We introduce a new analytical framework for the performance analysis of RIS-assisted systems.
Considering the impact of outdated CSI, we derive accurate closed-form expressions for the EC in the near- and far-field regions of the RISs.
In order to gain additional insights on the impact of the system parameters, we derive tight lower and upper bounds for the EC.
\item
Capitalizing on the obtained analytical results, we analyze the impact of key system and channel parameters on the
performance of RIS-aided systems, from which we conclude that the system performance improves with the transmit power,
the Rician-$K$ factor, the CSI accuracy and the size of the reflecting elements.
We also observe that the EC does not increase indefinitely when the number of reflecting elements tends to infinity.
Furthermore, the relative gains of the centralized and distributed deployments depend on the CSI accuracy and the location of the RISs.
The centralized deployment is shown to outperform the distributed deployment when the RIS is located near the BS or near the user,
while the distributed deployment usually offers better performance in the other scenarios.
\end{itemize}

The remainder of this paper is organized as follows. In Section II, we introduce the system and channel models.
In Section III, accurate closed-form approximate expressions for the EC are derived.
Moreover, tight upper and lower bounds for the EC are presented.
In Section IV, numerical and simulation results are illustrated to confirm the accuracy of the derived expressions.
Finally, Section V concludes the paper.

\newcounter{mytempeqncnt}

\section{System Model}
As illustrated in Fig. 1, we consider a SISO system in which a fixed single-antenna BS communicates with a single-antenna mobile user
with the assistance of $M$ passive reflecting elements, with each element being capable of independently adjusting its phase shift to
reflect the incident signals towards desired directions.
Two strategies for deploying the $M$ reflecting elements are considered: the centralized and distributed deployments.
As far as the centralized deployment is concerned, the $M$ reflecting elements are installed on one RIS.
As far as the distributed deployment is concerned, on the other hand, the $M$ elements are placed on $L$ ($L \ge 2$) RISs,
where the $l$-th RIS is equipped with ${M_l}$ reflecting elements.
It is worth noting that when the BS is sufficiently close to the RIS or the number of reflecting elements of the RIS is very large,
the far-field assumption, which means that the channel gain is the same for all the reflecting elements of the RIS,
does not hold anymore. In these cases, the RIS operates in the near-field region of the BS.
The boundary between the near- and far-field regions of the RIS is conventionally defined as
${D_{{\rm{boundary}}}} \buildrel \Delta \over = {{2D_{{\rm{RIS}}}^2} \mathord{\left/
 {\vphantom {{2D_{{\rm{RIS}}}^2} \lambda }} \right.
 \kern-\nulldelimiterspace} \lambda }$, where ${D_{{\rm{boundary}}}}$ is the distance between the BS and the center of the RIS,
 ${D_{{\rm{RIS}}}}$ is the maximum dimension of the RIS, and $\lambda $ is the wavelength of the signal.
 In the next sections, we introduce the near- and far-field system models for the two considered RIS deployment strategies.

\begin{figure}
  \centering
  \includegraphics[scale=0.6]{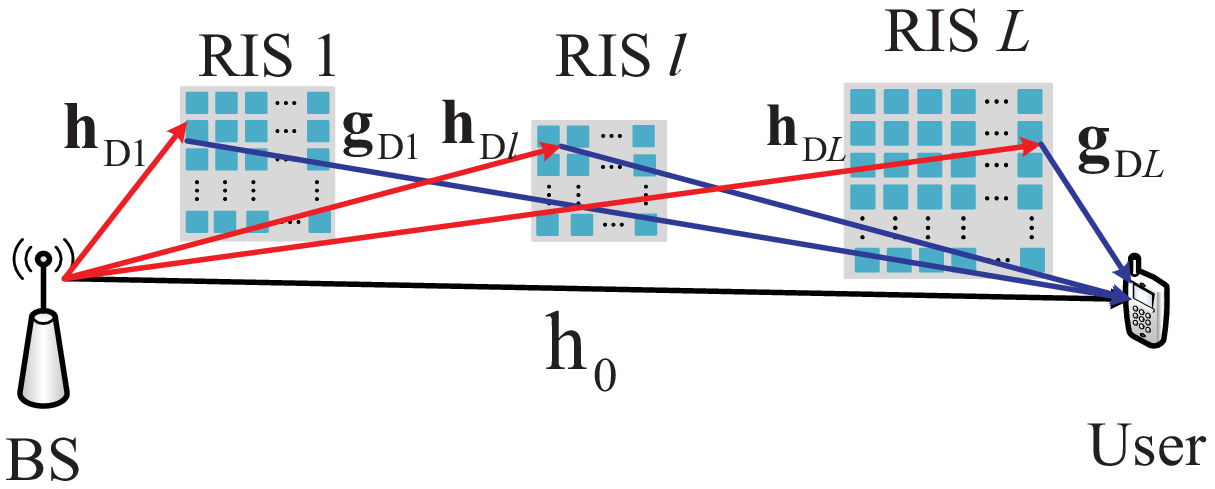}
  \hspace{0.3in}
  \includegraphics[scale=0.8]{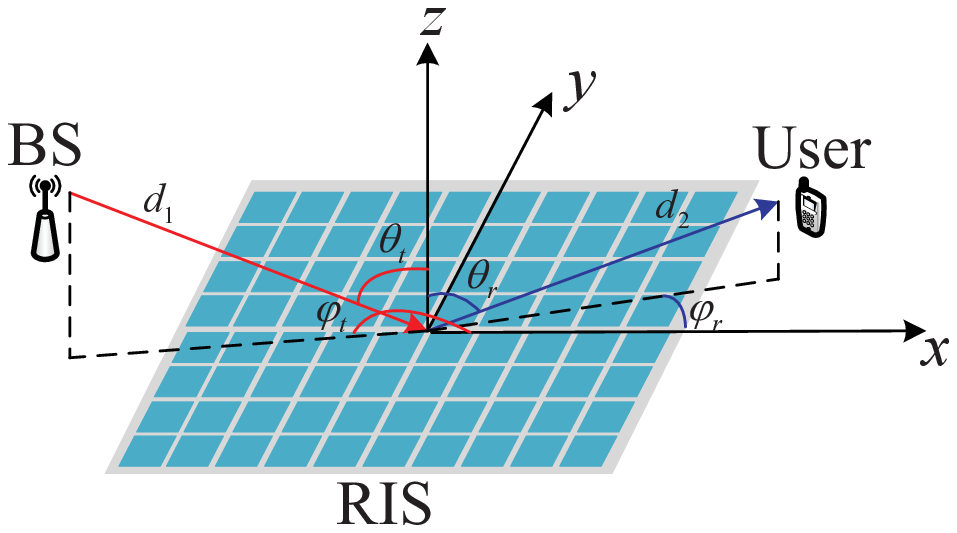}
  {\color{blue}\caption{RIS-aided communication system.}}
\end{figure}

\subsection{Centralized Deployment}
In the centralized deployment, we assume that the BS is located in $\left( {{x_{{\rm{t}}}},{y_{{\rm{t}}}},{z_{{\rm{t}}}}} \right)$,
the user is located in $\left( {{x_{\rm{r}}},{y_{\rm{r}}},{z_{\rm{r}}}} \right)$ and the center position of the RIS is
$\left( {{x_{\rm{0}}},{y_{\rm{0}}},{z_{\rm{0}}}} \right)$.
The RIS is deployed on a plane that is parallel to the $xy$-plane and is equipped with ${M_x} \times {M_y} = M$ reflecting elements.
The size of each element along the $x$ axis and the $y$ axis is ${d_x} $ and $ {d_y}$, respectively.
Therefore, the center position of the element of the RIS in the $y$-th row and $x$-th column is
$\left( {{x_0} + \left( {x - \frac{1}{2}} \right){d_x},{y_0} + \left( {y - \frac{1}{2}} \right){d_y},{z_0}} \right)$,
where $x \in \left[ {1 - {{{M_x}} \mathord{\left/
 {\vphantom {{{M_x}} 2}} \right.
 \kern-\nulldelimiterspace} 2},{{{M_x}} \mathord{\left/
 {\vphantom {{{M_x}} 2}} \right.
 \kern-\nulldelimiterspace} 2}} \right]$, $y \in \left[ {1 - {{{M_y}} \mathord{\left/
 {\vphantom {{{M_y}} 2}} \right.
 \kern-\nulldelimiterspace} 2},{{{M_y}} \mathord{\left/
 {\vphantom {{{M_y}} 2}} \right.
 \kern-\nulldelimiterspace} 2}} \right]$.

Due to the user mobility, it is not usually possible to perfectly estimate the CSI at the BS and the RIS
(e.g., due to the acquisition delay and the feedback overhead).
Therefore, we focus our attention on the impact of outdated CSI on the system performance.
We denote by ${{\rm{h}}_0}$ the direct channel from the BS to the user,
by ${\mathbf{h}_{\rm{C}}} = {\left[ {{{\rm{h}}_{{\rm{C}}1}}, \ldots ,{{\rm{h}}_{{\rm{C}}m}},
\ldots ,{{\rm{h}}_{{\rm{C}}M}}} \right]^T} \in {\mathbb{C}^{M }}$ the channel vector from the BS to the RIS,
and by  ${\mathbf{g}_{\rm{C}}} = \left[ {{{\rm{g}}_{{\rm{C}}1}}, \ldots ,{{\rm{g}}_{{\rm{C}}m}},
\ldots ,{{\rm{g}}_{{\rm{C}}M}}} \right] \in {\mathbb{C}^{1 \times M}}$  the channel vector from the RIS to the user.
More specifically, ${{\rm{h}}_0}$ and ${{\mathbf{g}_{\rm{C}}}}$ can be expressed, respectively, as \cite[Eq. (7)]{yang2020deepoutdated1},
\cite[Eq. (12)]{hashemi2021deep}
\begin{align}
\label{directlink}
{{\rm{h}}_0} = {\rho _0}{{{\rm{\hat h}}}_0} + {{\bar \rho }_0}{\omega _0},
\end{align}
\begin{align}
\label{RIS_D}
{{\bf{g}}_{\rm{C}}} = {\rho _{\rm{C}}}{{{\bf{\hat g}}}_{\rm{C}}} + {{\bar \rho }_{\rm{C}}}{\bm{\omega} _{\rm{C}}}.
\end{align}
where $0 \le {\rho _0} \le 1$ represents the correlation coefficient between the outdated channel estimate ${{\rm{\hat h}}_0}$ and
the actual channel ${{\rm{h}}_0}$, which can be calculated as
${\rho _0} = {J_0}\left( {2\pi {f_d}{T_{s0}}} \right)$ based on Clarke's fading spectrum \cite{yang2020deepoutdated1},
where ${J_0}\left( . \right)$ is the zeroth-order Bessel function of the first kind \cite[Eq. (8.411)]{Table}, ${f_d} = {{{f_c}v} \mathord{\left/
 {\vphantom {{{f_c}v} c}} \right.
\kern-\nulldelimiterspace} c}$ is the maximum Doppler shift, $f_c$ denotes the carrier frequency,
$v$ denotes the velocity of the user, $c$ denotes the speed of light,
and ${{T_{s0}}}$ is the estimation delay between the actual channel and the outdated channel. In addition, ${{\bar \rho }_0} \buildrel \Delta \over =  {\sqrt {1 - \rho _0^2} } $.
Similarly, ${\rho _\mathrm{C}} = {J_0}\left( {2\pi {f_d}{T_{sC}}} \right)$ is the correlation coefficient between the outdated channel
${\mathbf{{\hat g}}_{\rm{C}}} = \left[ {{{\mathbf{{\rm{\hat g}}}}_{{\rm{C1}}}}, \ldots ,{{{\rm{\hat g}}}_{{\rm{C}}m}},
\ldots ,{{{\rm{\hat g}}}_{{\rm{C}}M}}} \right] \in {\mathbb{C}^{1 \times M}}$ and the actual channel ${{\mathbf{g}_{\rm{C}}}}$, and ${{\bar \rho }_{\rm{C}}} \buildrel \Delta \over =  {\sqrt {1 - \rho _{\rm{C}}^2} } $.
If ${\rho _i} = 1$ $\left( {i \in \left\{ {0,{\rm{C}}} \right\}} \right)$, the CSI is perfect,
whereas ${\rho _i} = 0$ indicates no CSI.
In addition, ${\omega _0} \sim {\cal C}{\cal N}\left( {0,\sigma _{{{{\rm{\hat h}}}_0}}^2} \right)$, where ${\cal C}{\cal N}\left( {0,{\sigma ^2}} \right)$ is the complex Gaussian distribution with zero mean and variance $\sigma ^2$, ${\bm{\omega} _{\rm{C}}}
= \left[ {{\omega _{{\rm{C1}}}}, \ldots ,{\omega _{{\rm{C}}m}}, \ldots ,{\omega _{{\rm{C}}M}}} \right] \in {\mathbb{C}^{1 \times M}}$
with ${\omega _{{\rm{C}}m}} \sim {\cal C}{\cal N}\left( {0,\sigma _{{\mathrm{{\hat g}}_{{\rm{C}}m}}}^2} \right)\left( {m = 1, \ldots ,M} \right)$.
We assume that all links experience Rician\footnote{It is worth mentioning that the method that we utilize to obtain the derivations of the performance of RIS-assisted systems can be readily applied to other channel fading models (e.g., Nakagami-$m$ \cite{tegos2021distribution, yang2012multiuser}, $\kappa- \mu $ \cite{kumar2014analysis}). The analysis of different small scale fading models in the near- and far-field regions of RIS-assisted systems is postponed to a future research work.} fading,
i.e., ${{{\rm{\hat h}}}_0} = \sqrt {\frac{{{K_0}}}{{1 + {K_0}}}} {\rm{\hat h}}_{\rm{0}}^{{\rm{LoS}}} + \sqrt {\frac{1}{{1 + {K_0}}}}
{\rm{\hat h}}_{\rm{0}}^{{\rm{NLoS}}}$,
${\mathbf{h}_{\rm{C}}} = \sqrt {\frac{{{K_1}}}{{1 + {K_1}}}}\mathbf{ h}_{\rm{C}}^{{\rm{LoS}}}
+ \sqrt {\frac{1}{{1 + {K_1}}}} \mathbf{h}_{\rm{C}}^{{\rm{NLoS}}}$
and ${\mathbf{{\hat g}}_{\rm{C}}} = \sqrt {\frac{{{K_2}}}{{1 + {K_2}}}} \mathbf{\hat g}_{\rm{C}}^{{\rm{LoS}}} +
 \sqrt {\frac{1}{{1 + {K_2}}}} \mathbf{\hat g}_{\rm{C}}^{{\rm{NLoS}}}$,
where ${K_i}\left( {i \in \left\{0,1,2\right\}} \right)$ is the Rician-$K$ factor,
${{\rm{X}}^{{\rm{LoS}}}}$  $\left( {{\rm{X  }}\in \left\{ {{{{\rm{\hat h}}}_0},
{{{\bf{\hat h}}}_\mathrm{C}}, {{{\bf{\hat g}}}_\mathrm{C}}} \right\}} \right)$
denotes the line-of-sight (LoS) component,
and ${{\rm{X}}^{{\rm{NLoS}}}}$ denotes the non-LoS (NLoS) component.
It is worth noting that there exist several efficient channel estimation methods for RIS-assisted communication systems, such as the minimum mean square error \cite{ alwazani2020intelligent } and deep learning \cite{ kundu2021channel } methods. In this paper, the channel from the BS to the RIS is assumed to be perfectly estimated because the BS and the RIS are assumed to be at fixed locations \cite{yu2020robustoutdated2}.
As a result, the received signal at the user can be written as
\begin{align}
 \label{received_signal}
&{y_{\rm{C}}} \!=\!\! \sqrt P\! \left(\! {{{\bf{g}}_{\rm{C}}}{\bf{B}}{{\bf{\Phi }}_{\rm{C}}}{{\bf{h}}_{\rm{C}}} \!+\! \sqrt {\beta _0^{ - 1}} {{\rm{h}}_0}} \!\right)\!s\! +\! {n_0} \!\nonumber\\
&=\! \sqrt P \!\left(\! {\sum\limits_{m = 1}^M {\frac{1}{{\sqrt {{\beta _m}} }}{\mathrm{g}_{{\rm{C}}m}}{\mathrm{h}_{{\rm{C}}m}}{e^{j{\varphi _{{\rm{C}}m}}}}}  + \frac{1}{{\sqrt {\beta _0^{ - 1}} }}{{\rm{h}}_0}} \!\right)\!\!s \!+\! {n_0},
 \end{align}
where ${{\rm{g}}_0} \buildrel \Delta \over = \sqrt {\beta _0^{ - 1}}{{\rm{h}}_0}$, $P$ is the transmit power, $n_0$ is the zero-mean additive white Gaussian noise (AWGN) whose variance is $\sigma _0^2$,
${{\beta _0}}$ is the path loss of the direct link,
${\bm{\Phi} _{\rm{C}}} = \mathrm{diag}\left\{ {{e^{j{\varphi _{{\rm{C1}}}}}}, \ldots ,{e^{j{\varphi _{{\rm{C}}M}}}}} \right\} $
where ${\varphi _{{\rm{C}}m}}\left( {m = 1, \ldots ,M} \right)$ is the phase shift of the $m$-th element of the RIS, and
$s$ is the transmit signal with unit energy.
Moreover, $\bm{{\rm B}}  \buildrel \Delta \over = {\rm{diag}}\left\{ {{{{\left( {\sqrt {{\beta _1}} } \right)}^{ - 1}}} ,
\ldots ,{{{\left( {\sqrt {{\beta _M}} } \right)}^{ - 1}}} } \right\}$ represents the path loss matrix
with ${\beta _m} \left( {m = 1, \ldots M} \right)$ denoting the path loss of the $m$-th element of the RIS\footnote{The generalization of the proposed analytical framework in the presence of channel correlation \cite{gradoni2021end} (i.e., $\bm{{\rm B}}$ is a non-diagonal matrix) is left for future research.}, which,
according to \cite{tang2021path}, and under the assumption that the peak radiation directions of the transmitting and receiving antennas
point towards the center of the RIS, can be expressed as
\begin{align}
\label{beta_m}
{\beta _m} \buildrel \Delta \over = {{{\beta _0}{{\left( {r_m^{\rm{t}}r_m^{\rm{r}}} \right)}^2}} \mathord{\left/
 {\vphantom {{{\beta _0}{{\left( {r_m^{\rm{t}}r_m^{\rm{r}}} \right)}^2}} {F_m^{{\rm{combine}}}}}} \right.
 \kern-\nulldelimiterspace} {F_m^{{\rm{combine}}}}}\;,\;\;m = 1,2, \ldots ,M,
\end{align}
\noindent where ${\beta _0} \buildrel \Delta \over = {{16{\pi ^2}} \mathord{\left/
 {\vphantom {{16{\pi ^2}} {\left( {{G_{\rm{t}}}{G_{\rm{r}}}d_x^2d_y^2} \right)}}} \right.
 \kern-\nulldelimiterspace} {\left( {{G_{\rm{t}}}{G_{\rm{r}}}d_x^2d_y^2} \right)}}$, ${{G_{\rm{t}}}}$ and ${{G_{\rm{r}}}}$ represent the transmit antenna gain and the receive antenna gain,
respectively, and ${r_m^{\rm{t}}}$ and ${r_m^{\rm{r}}}$ denote the distance between the BS and the $m$-th element of the
RIS and the distance between the $m$-th element of the RIS and the user, respectively.
Furthermore, $F_m^{{\rm{combine}}}$ is the joint normalized power radiation pattern,
which depends on the location of the BS, the RIS, and the user, and is defined as
\begin{align}
\label{Fcombine}
\!F_m^{{\rm{combine}}} \!\!\buildrel \Delta \over = \!\!{\left( {\cos \theta _m^{{\rm{tx}}}} \right)^{\frac{{{G_{\rm{t}}}}}{2} - 1}}
\!\cos\! \left( {\theta _m^{\rm{t}}} \right)\!\cos\! \left( {\theta _m^{\rm{r}}} \right)\!
{\left( {\cos\! \theta _m^{{\rm{rx}}}} \right)^{\frac{{{G_{\rm{r}}}}}{2} - 1}},
\end{align}
where ${\theta _m^{{\rm{tx}}}}$ and ${\theta _m^{{\rm{rx}}}}$ denote the angles of elevation from the BS antenna and the
user antenna to the $m$-th reflecting element of the RIS, respectively.
 In addition, ${\theta _m^{\rm{t}}}$ and ${\theta _m^{\rm{r}}}$ represent the angles of elevation from the $m$-th reflecting
 element of the RIS to the BS antenna and user antenna, respectively. According to \cite{tang2021path}, we have
$\cos \theta _m^{{\rm{tx}}} = {{\left( {d_1^2 + {{\left( {r_m^{\rm{t}}} \right)}^2} - d_m^2} \right)} \mathord{\left/
 {\vphantom {{\left( {d_1^2 + {{\left( {r_m^{\rm{t}}} \right)}^2} - d_m^2} \right)} {\left( {2{d_1}r_m^t} \right)}}} \right.
 \kern-\nulldelimiterspace} {\left( {2{d_1}r_m^t} \right)}}$,
 $\cos \theta _m^{{\rm{rx}}} = {{\left( {d_2^2 + {{\left( {r_m^{\rm{r}}} \right)}^2} - d_m^2} \right)} \mathord{\left/
 {\vphantom {{\left( {d_2^2 + {{\left( {r_m^{\rm{r}}} \right)}^2} - d_m^2} \right)} {\left( {2{d_2}r_m^{\rm{r}}} \right)}}} \right.
 \kern-\nulldelimiterspace} {\left( {2{d_2}r_m^{\rm{r}}} \right)}}$,
 $\cos \theta _m^{\rm{t}} = {{\left( {{z_{\rm{t}}} - {z_0}} \right)} \mathord{\left/
 {\vphantom {{\left( {{z_{\rm{t}}} - {z_0}} \right)} {r_m^{\rm{t}}}}} \right.
 \kern-\nulldelimiterspace} {r_m^{\rm{t}}}}$
 and $\cos \theta _m^{\rm{r}} = {{\left( {{z_{\rm{t}}} - {z_0}} \right)} \mathord{\left/
 {\vphantom {{\left( {{z_{\rm{t}}} - {z_0}} \right)} {r_m^{\rm{r}}}}} \right.
 \kern-\nulldelimiterspace} {r_m^{\rm{r}}}}$,
where $d_1$,  $d_2$ and $d_m$ denote the distance between the BS and the center of the RIS,
the distance between the user and the center of the RIS,
and the distance between the $m$-th element of the RIS and the center of the RIS, respectively.
Further details on the path loss model for RIS-aided wireless communications can be found in \cite{tang2021path}.

\noindent By substituting \eqref{directlink} and \eqref{RIS_D} into \eqref{received_signal}, we can rewrite the received signal as
\begin{align}
\label{received_signal_eff}
&{y_{\rm{C}}} = \underbrace {\sqrt P \left( {{\rho _{\rm{C}}}{{{\bf{\hat g}}}_{\rm{C}}}{\bf{B}}{{\bf{\Phi }}_{\rm{C}}}{{\bf{h}}_{\rm{C}}} +  {\rho _0}{{{\rm{\hat g}}}_0}} \right)s}_{{\rm{desired }}\;{\rm{signal}}}\nonumber\\
 &+ \underbrace {\underbrace {\sqrt P \left( {{{\bar \rho }_{\rm{C}}} {{\bm{\omega }}_{\rm{C}}}{\bf{B}}{{\bf{\Phi }}_{\rm{C}}}{{\bf{h}}_{\rm{C}}} + \sqrt {\beta _0^{ - 1}} {{\bar \rho }_{\rm{0}}}{\omega _0}} \right)s}_{{\rm{outdated}}\;{\rm{CSI}}\;{\rm{noise}}} + \underbrace {{n_0}}_{{\rm{white}}\;{\rm{noise}}}}_{{\rm{effective}}\;{\rm{noise}}\;{n_{{\rm{Ceff}}}}}.
\end{align}
where ${{{\rm{\hat g}}}_0} \buildrel \Delta \over = \sqrt {\beta _0^{ - 1}} {{{\rm{\hat h}}}_0}$. As can be seen from (6), the second term is the outdated CSI noise. Therefore, the effective noise ${{n_{{\rm{Ceff}}}}}$ is composed of the outdated CSI noise and the AWGN, so that the effective transmit SNR can be formulated as ${\gamma _{{\rm{Cteff}}}} \buildrel \Delta \over = {P \mathord{\left/
 {\vphantom {P {E\left( {{{\left| {{n_{{\rm{Ceff}}}}} \right|}^2}} \right)}}} \right.
 \kern-\nulldelimiterspace} {\mathbb{E}\left( {{{\left| {{n_{{\rm{Ceff}}}}} \right|}^2}} \right)}}$, where ${\mathbb{E}\left( {{{\left| {{n_{{\rm{Ceff}}}}} \right|}^2}} \right)}$ is calculated in the next section. Hence, the received SNR at the user can be formulated by using \eqref{received_signal_eff} as
\begin{align}
\label{SNR_cen}
{\gamma _{{\rm{C}}}} = {\gamma _{{\rm{Cteff}}}}{{{\left| {{\rho _{\rm{C}}}{\mathbf{{\hat g}}_{\rm{C}}}
\bm{{\rm B}}{\bm{\Phi} _{\rm{C}}}{\mathbf{h}_{\rm{C}}} + {\rho _0}{{{\rm{\hat g}}}_0}} \right|}^2}},
\end{align}
The phase shifts at the RIS can be designed such that the outdated direct channel (from the BS to the user) and the outdated cascaded channel (from the BS to the RIS and from the RIS to the user) are co-phased, i.e., ${\varphi _{{\rm{C}}m}} = {\varphi _{{{{\rm{\hat h}}}_0}}} - \left( {{\varphi _{{\mathrm{{\hat g}}_{{\rm{C}}m}}}} + {\varphi _{{\mathrm{h}_{{\rm{C}}m}}}}} \right)$ for $ {m = 1, \ldots ,M}  $ \cite{galappaththige2020performance}, and, therefore,
the signals via the two channels are constructively added at the user and the received SNR is maximized. In this case, we obtain
\begin{align}
\label{SNR_cen_max}
\!{\gamma _{{\rm{Cmax}}}} \!= \!{\gamma _{{\rm{Cteff}}}}\!{\left( \!{{\rho _{\rm{C}}}\!\sum\limits_{m = 1}^M\! {\sqrt {\beta _m^{ - 1}} \!\left| {{{{\rm{\hat g}}}_{{\rm{C}}m}}} \right|\!\left| {{{\rm{h}}_{{\rm{C}}m}}} \right| \!+ \!{\rho _0}\left| {{{{\rm{\hat g}}}_0}} \right|} } \!\right)^2}.
\end{align}

{\bf{Far-field case:}}
The channel in \eqref{beta_m} is a general path loss model that can be applied in the near-field and far-field regions of the RIS.
In the far-field case, \eqref{beta_m} can be simplified since we have
${d_m} \ll r_m^{\rm{t}} \approx {d_1}$ and ${d_m} \ll r_m^{\rm{r}} \approx {d_2}$,
which results in $\cos \theta _m^{{\rm{tx}}} \approx 1$ and $\cos \theta _m^{{\rm{rx}}} \approx 1$.
Therefore, \eqref{Fcombine} reduces to $\cos \left( {{\theta ^\mathrm{t}}} \right)\cos \left( {{\theta ^\mathrm{r}}} \right)$
and \eqref{beta_m} can be simplified as
\begin{align}
\label{betam_far}
\!{\beta ^{{\rm{farfield}}}} \!\buildrel \Delta \over = \!{{{\beta _0}{{\left( {{d_1}{d_2}} \right)}^2}}\!\! \mathord{\left/
 {\vphantom {{{\beta _0}{{\left( {{d_1}{d_2}} \right)}^2}}\!\! {\left(\! {\cos\! \left( {{\theta ^{\rm{t}}}} \right)\!\cos \! \left( {{\theta ^{\rm{r}}}} \right)} \right)}}} \right.
 \kern-\nulldelimiterspace}\!\! {\left( {\cos\! \left(\! {{\theta ^{\rm{t}}}}\! \right)\!\cos\! \left(\! {{\theta ^{\rm{r}}}} \!\right)} \!\right)}}\!\;,\; m = 1,2, \ldots ,M,
\end{align}
where $\cos \left( {{\theta ^\mathrm{t}}} \right) = {{\left( {{z_{\rm{t}}} - {z_0}} \right)} \mathord{\left/
 {\vphantom {{\left( {{z_{\rm{t}}} - {z_0}} \right)} {{d_1}}}} \right.
 \kern-\nulldelimiterspace} {{d_1}}}$,
 $\cos \left( {{\theta ^\mathrm{r}}} \right) = {{\left( {{z_{\rm{r}}} - {z_0}} \right)} \mathord{\left/
 {\vphantom {{\left( {{z_{\rm{r}}} - {z_0}} \right)} {{d_2}}}} \right.
 \kern-\nulldelimiterspace} {{d_2}}}$.
 From \eqref{betam_far}, we see that all the elements of the RIS have the same path loss.
 When the BS and the user are in the far-field of the RIS, therefore, the received signal can be simplified as
\begin{align}
\label{SNR_cen_max_far}
{y_{\rm{C}}} &= \sqrt P \left( {{\rho _{\rm{C}}}{{\left( {\sqrt {{\beta ^{{\rm{farfield}}}}} } \right)}^{ - 1}}{{{\bf{\hat g}}}_{\rm{C}}}{\mathbf{\Phi} _{\rm{C}}}{{\bf{h}}_{\rm{C}}} + {\rho _0}{{{\rm{\hat g}}}_0}} \right)\nonumber\\
& + \underbrace {\sqrt P {{\bar \rho }_{\rm{C}}}{{\left( {\sqrt {{\beta ^{{\rm{farfield}}}}} } \right)}^{ - 1}}{\bm{\omega} _{\rm{C}}}{\mathbf{\Phi} _{\rm{C}}}{{\bf{h}}_{\rm{C}}}s + {\omega _e}}_{{n_{{\rm{Ceff}}}}}.
\end{align}
where ${\omega _e} \buildrel \Delta \over = \sqrt {P\beta _0^{ - 1}} {{\bar \rho }_{\rm{0}}}{\omega _0}s + {n_0}$.

\noindent By employing the same phase shift design as that in \eqref{SNR_cen_max}, the optimal received SNR is
\begin{align}
\!\!\gamma _{{\rm{Cmax}}}^{{\rm{farfield}}} \!=\! \gamma _{{\rm{Cteff}}}^{{\rm{farfield}}}\!{\left(\!\! {{\rho _{\rm{C}}}\!{{\left(\!\! {\sqrt {{\beta ^{{\rm{farfield}}}}} } \right)}^{ - 1}}\!\!\sum\limits_{m = 1}^M \! \left| {{{{\rm{\hat g}}}_{{\rm{C}}m}}} \right|\!\left| {{{\rm{h}}_{{\rm{C}}m}}} \right| \!+ \!{\rho _0}\!\left| {{{{\rm{\hat g}}}_0}} \right|} \!\right)^2}\!\!,
\end{align}
where $\gamma _{{\rm{Cteff}}}^{{\rm{far  field}}} \buildrel \Delta \over = {P\!\! \mathord{\left/
 {\vphantom {P {E\left( {{{\left| {n_{{\rm{Ceff}}}^{{\rm{far  field}}}} \right|}^2}} \right)}}} \right.
 \kern-\nulldelimiterspace} {\mathbb{E}\left( {{{\left| {n_{{\rm{Ceff}}}^{{\rm{far  field}}}} \right|}^2}} \right)}}.$

\subsection{Distributed Deployment}
In the distributed deployment, we assume that the BS and the user are located in the same positions as those of the
centralized deployment.
The center position of the $l$-th RIS that comprises ${M_{xl}} \times {M_{yl}} = {M_l}$ reflecting
elements is $\left( {{x_{0l}},{y_{0l}},{z_{0l}}} \right)$ and all the RISs are deployed parallel to the $xy$-plane.
Note that all the RISs can be intelligently controlled to reflect the incident signals towards the user so that there is no interference \cite{abrardo2021mimo} among them \cite{renzo2020smart, yang2020outage,do2021multi,galappaththige2020performance,arun2020rfocus}.
Similar to the centralized deployment, all RIS elements have the same size ${d_x} \times {d_y}$.
Thus, the center position of the element in the $y$-th row and $x$-th column of the $l$-th RIS is
$\left( {{x_{0l}} + \left( {x - \frac{1}{2}} \right){d_x},{y_{0l}} + \left( {y - \frac{1}{2}} \right){d_y},{z_{0l}}} \right)$,
where $x \in \left[ {1 - {{{M_{xl}}} \mathord{\left/
 {\vphantom {{{M_{xl}}} 2}} \right.
 \kern-\nulldelimiterspace} 2},{{{M_{xl}}} \mathord{\left/
 {\vphantom {{{M_{xl}}} 2}} \right.
 \kern-\nulldelimiterspace} 2}} \right],y \in \left[ {1 - {{{M_{yl}}} \mathord{\left/
 {\vphantom {{{M_{yl}}} 2}} \right.
 \kern-\nulldelimiterspace} 2},{{{M_{yl}}} \mathord{\left/
 {\vphantom {{{M_{yl}}} 2}} \right.
 \kern-\nulldelimiterspace} 2}} \right]$.
Let ${\mathbf{h}_{{\rm{D}}l}} = {\left[ {{{\rm{h}}_{{\rm{D}}l1}}, \ldots ,{{\rm{h}}_{{\rm{D}}lm}},
\ldots ,{{\rm{h}}_{{\rm{D}}l{M_l}}}} \right]^T} \in {\mathbb{C}^{{M_l}}}$
and ${\mathbf{g}_{{\rm{D}}l}} = \left[ {{{\rm{g}}_{{\rm{D}}l1}}, \ldots ,{{\rm{g}}_{{\rm{D}}lm}},
\ldots ,{{\rm{g}}_{{\rm{D}}l{M_l}}}} \right] \in {\mathbb{C}^{1 \times {M_l}}}$
denote the channel vectors between the BS and the RIS, and between the RIS and the user, respectively.
In the presence of outdated CSI, ${\mathbf{g}_{{\rm{D}}l}}$ can be written as
 \begin{align}
 \label{g_Dl}
 {{\bf{g}}_{{\rm{D}}l}} = {\rho _{{\rm{D}}l}}{{{\bf{\hat g}}}_{{\rm{D}}l}} + {{\bar \rho }_{{\rm{D}}l}}{\mathbf{\omega} _{{\rm{D}}l}},
 \end{align}
 where ${\mathbf{{\hat g}}_{{\rm{D}}l}} = \left[ {{{{\rm{\hat g}}}_{{\rm{D}}l1}}, \ldots ,
 {{{\rm{\hat g}}}_{{\rm{D}}lm}}, \ldots ,{{{\rm{\hat g}}}_{{\rm{D}}l{M_l}}}} \right] \in {\mathbb{C}^{1 \times {M_l}}}$,
 ${\rho _{{\rm{D}}l}}$ is the correlation coefficient between ${\mathbf{{\hat g}}_{{\rm{D}}l}} $ and ${\mathbf{g}_{{\rm{D}}l}}$, and ${{\bar \rho }_{{\rm{D}}l}} \buildrel \Delta \over = {\sqrt {1 - \rho _{{\rm{D}}l}^2} } $.
${{\bm{\omega }}_{{\rm{D}}l}} = \left[ {{\omega _{{\rm{D}}l1}}, \ldots ,{\omega _{{\rm{D}}lm}}, \ldots ,
{\omega _{{\rm{D}}l{M_l}}}} \right] \in {\mathbb{C}^{1 \times {M_l}}}$
 with ${\omega _{{\rm{D}}lm}} \sim {\cal C}{\cal N}\left( {0,\sigma _{{\mathrm{{\hat g}}_{{\rm{D}}lm}}}^2} \right)
 \left( {m = 1, \ldots ,{M_l}} \right)$.
 In addition, all the links experience Rician fading, from which we have
 ${\mathbf{h}_{{\rm{D}}l}} = \sqrt {\frac{{{K_{1l}}}}{{1 + {K_{1l}}}}} \mathbf{h}_{{\rm{D}}l}^{{\rm{LoS}}}
 + \sqrt {\frac{1}{{1 + {K_{1l}}}}} \mathbf{h}_{{\rm{D}}l}^{{\rm{NLoS}}}$
 and ${\mathbf{{\hat g}}_{{\rm{D}}l}} = \sqrt {\frac{{{K_{2l}}}}{{1 + {K_{2l}}}}} \mathbf{\hat g}_{{\rm{D}}l}^{{\rm{LoS}}}
 + \sqrt {\frac{1}{{1 + {K_{2l}}}}} \mathbf{\hat g}_{{\rm{D}}l}^{{\rm{NLoS}}}$,
 where $K_{1l}$ and $K_{2l}$ are the Rician-$K$ factors of ${\mathbf{h}_{{\rm{D}}l}}$
 and ${\mathbf{{\hat g}}_{{\rm{D}}l}}$, respectively,
 ${{\rm{Y}}^{{\rm{LoS}}}}$  $\left( {{\rm{Y }} \in \left\{ {{{{\bf{\hat h}}}_{{\rm{D}}l}},
 {{{\bf{\hat g}}}_{{\rm{D}}l}}} \right\}} \right)$ is the LoS component,
 and ${{\rm{Y}}^{{\rm{NLoS}}}}$ is the  NLoS component.
 Then, the received signal at the user is
 \begin{align}
 \label{received_signal_dis}
 \!{y_{\rm{D}}} \!=\! \sqrt P \left( {\sum\limits_{l = 1}^L {{\mathbf{g}_{{\rm{D}}l}}{\bm{{\rm B}}_l}{\bm{\Phi}
 _{{\rm{D}}l}}{\mathbf{h}_{{\rm{D}}l}}}  \!+\! {{\rm{g}}_0}} \right)s \!+\! {n_0},
 \end{align}
 where ${\bm{\Phi} _{{\rm{D}}l}} \!=\!\mathrm{diag}\!\!\left\{\! {{e^{j{\varphi _{{\rm{D}}l{\rm{1}}}}}},\!
 \ldots \!,{e^{j{\varphi _{{\rm{D}}l{M_l}}}}}} \!\right\}$ denotes the reflection coefficient of the $l$-th RIS,
 ${\bm{{\rm B}}_l}\! \buildrel \Delta \over =\! {\rm{diag}}\!\!
 \left\{\!\! {{{\left(\! {\sqrt {{\!\beta _{l1}}} } \right)}^{ - 1}}\!,
 \ldots ,\!{{\left(\! {\sqrt {{\!\beta _{l{M_l}}}} }\! \right)}^{ - 1}}} \!\right\}$
 is the path loss matrix with ${\beta _{lm}}\left( {m = 1, \ldots ,{M_l},l = 1, \ldots ,L} \right)$
 denoting the path loss of the $m$-th element of the $l$-th RIS, which is given by
 \begin{align}
\!\!\!\!\!{\beta _{lm}}{\rm{ }} \!\buildrel \Delta \over = \!{{{\beta _0}\!{{\left( {r_{lm}^{\rm{t}}r_{lm}^{\rm{r}}} \right)}^2}} \!\! \mathord{\left/
 {\vphantom {{{\beta _0}\!{{\left( {r_{lm}^{\rm{t}}r_{lm}^{\rm{r}}} \right)}^2}} \!\!{F_{lm}^{{\rm{combine}}}}}} \right.
 \kern-\nulldelimiterspace}\!\! {F_{lm}^{{\rm{combine}}}}}\!\!,\!\;m \!=\! 1,2,\! \ldots \!,M,l \!=\! 1,\! \ldots \!,L,
 \end{align}
 with
\begin{align}
\!\!\!\!F_{lm}^{{\rm{combine}}} \!\buildrel \Delta \over = \!\!{\left(\! {\cos\! \theta _{lm}^{{\rm{tx}}}} \right)^{\frac{{{G_{\rm{t}}}}}{2} - 1}}
\!\cos \!\left(\! {\theta _{lm}^{\rm{t}}} \right)\!\cos\! \left( {\theta _{lm}^{\rm{r}}} \right)
\!{\left(\! {\cos\! \theta _{lm}^{{\rm{rx}}}} \right)^{\frac{{{G_{\rm{r}}}}}{2} - 1}}\!,
\end{align}
where ${\theta _{lm}^{{\rm{tx}}}}$, ${\theta _{lm}^{{\rm{rx}}}}$, ${\theta _{lm}^{\rm{t}}}$
and ${\theta _{lm}^{\rm{r}}}$ represent the angle of elevation from the BS antenna to the $m$-th reflecting element of the $l$-th RIS,
the angle of elevation from the user antenna to the $m$-th reflecting element of the $l$-th RIS,
the angle of elevation from the $m$-th reflecting element of the $l$-th RIS to the BS antenna,
and the angle of elevation from the $m$-th reflecting element of the $l$-th RIS to the user antenna, respectively.
From \cite{tang2021path}, we can write $\cos \theta _{lm}^{{\rm{tx}}} = {{\left( {d_{1l}^2
+ {{\left( {r_{lm}^{\rm{t}}} \right)}^2} - d_{lm}^2} \right)} \mathord{\left/
 {\vphantom {{\left( {d_{1l}^2 + {{\left( {r_{lm}^{\rm{t}}} \right)}^2}
 - d_{lm}^2} \right)} {\left( {2{d_{l1}}r_{lm}^{\rm{t}}} \right)}}} \right.
 \kern-\nulldelimiterspace} {\left( {2{d_{l1}}r_{lm}^{\rm{t}}} \right)}}$, $\cos \theta _{lm}^{{\rm{rx}}}
 = {{\left( {d_{2l}^2
 + {{\left( {r_{lm}^{\rm{r}}} \right)}^2} - d_{lm}^2} \right)} \mathord{\left/
 {\vphantom {{\left( {d_{2l}^2 + {{\left( {r_{lm}^{\rm{r}}} \right)}^2} - d_{lm}^2} \right)}
 {\left( {2{d_{2l}}r_{lm}^{\rm{r}}} \right)}}} \right.
 \kern-\nulldelimiterspace} {\left( {2{d_{2l}}r_{lm}^{\rm{r}}} \right)}}$,
$\cos \theta _{lm}^{\rm{t}} = {{\left( {{z_{\rm{t}}} - {z_{0l}}} \right)} \mathord{\left/
 {\vphantom {{\left( {{z_{\rm{t}}} - {z_{0l}}} \right)} {r_{lm}^{\rm{t}}}}} \right.
 \kern-\nulldelimiterspace} {r_{lm}^{\rm{t}}}}$ and
 $\cos \theta _{lm}^{\rm{r}} = {{\left( {{z_{\rm{r}}} - {z_{0l}}} \right)} \mathord{\left/
 {\vphantom {{\left( {{z_{\rm{r}}} - {z_{0l}}} \right)} {r_{lm}^{\rm{r}}}}} \right.
 \kern-\nulldelimiterspace} {r_{lm}^{\rm{r}}}}$,
where $d_{1l}$, $d_{2l}$, ${r_{lm}^{\rm{t}}}$, ${r_{lm}^{\rm{r}}}$ and ${d_{lm}}$ represent the distance between
the BS and the $l$-th RIS, the distance between the user and the $l$-th RIS,
the distance between the BS and the $m$-th element of the $l$-th RIS,
the distance between the user and the $m$-th element of the $l$-th RIS,
and the distance between the $m$-th element of the $l$-th RIS to the center of the $l$-th RIS, respectively.
Substituting \eqref{g_Dl} and \eqref{directlink} into \eqref{received_signal_dis}, the received signal can be rewritten as
\begin{align}
\label{received_signal_dis_general}
{y_{\rm{D}}}\! & =\! \sqrt P\! \left( {\sum\limits_{l = 1}^L {{\rho _{{\rm{D}}l}}{{{\bf{\hat g}}}_{{\rm{D}}l}}{{\bf{B}}_{{\rm{D}}l}}{{\bf{\Phi }}_{{\rm{D}}l}}{{\bf{h}}_{{\rm{D}}l}}}  + {\rho _0}{{{\rm{\hat g}}}_{\rm{0}}}} \right)\!s \!\nonumber\\
&+\! \underbrace {\sqrt P \sum\limits_{l = 1}^L  {{\bar \rho }_{{\rm{D}}l}}{{\bf{\omega }}_{{\rm{D}}l}}{{\bf{B}}_{{\rm{D}}l}}{{\bf{\Phi }}_{{\rm{D}}l}}{{\bf{h}}_{{\rm{D}}l}}s + {\omega _e}}_{{n_{{\rm{Deff}}}}}.
 \end{align}
Therefore, the received SNR can be expressed as
\begin{align}
\gamma_{\rm{D}} = {\gamma _{{\rm{Dteff}}}}{\left| {\sum\limits_{l = 1}^L
 {{\rho _{{\rm{D}}l}}{\mathbf{{\hat g}}_{{\rm{D}}l}}{\bm{{\rm B}}_{{\rm{D}}l}}{\bm{\Phi} _{{\rm{D}}l}}{\mathbf{h}_{{\rm{D}}l}}}
  +   {\rho _0}{{{\rm{\hat g}}}_{\rm{0}}}} \right|^2},
\end{align}
where ${\gamma _{{\rm{Dteff}}}} \buildrel \Delta \over = {P\!\! \mathord{\left/
 {\vphantom {P {E\left( {{{\left| {{n_{{\rm{Deff}}}}} \right|}^2}} \right)}}} \right.
 \kern-\nulldelimiterspace}\! {\mathbb{E}\left( {{{\left| {{n_{{\rm{Deff}}}}} \right|}^2}} \right)}}$ is the effective SNR.
The optimal phase-shift matrix that maximizes the received SNR at the user can be expressed
as \cite{galappaththige2020performance} ${\varphi _{lm}} = \mathop {\arg \max }\limits_{ - \pi  \le {\varphi _m} \le \pi }
\left( {{\theta _{{{{\rm{\hat h}}}_{\rm{0}}}}} - \left( {{\theta _{{\mathrm{{\hat g}}_{{\rm{D}}lm}}}}
+ {\theta _{{\mathrm{h}_{{\rm{D}}lm}}}}} \right)} \right)$, for $m = 1, \ldots {M_l}$ and $l = 1, \ldots L$,
where ${{\theta _{{{{\rm{\hat h}}}_{\rm{0}}}}}}$, ${{\theta _{{\mathrm{{\hat g}}_{{\rm{D}}lm}}}}}$
and ${{\theta _{{\mathrm{h}_{{\rm{D}}lm}}}}}$ are the phases of ${{{{\rm{\hat h}}}_{\rm{0}}}}$,
${{\mathrm{{\hat g}}_{\mathrm{D}lm}}}$ and ${{\mathrm{h}_{\mathrm{D}lm}}}$, respectively. Hence, the maximum achievable SNR is
\begin{align}
\label{SNR_max_dis}
\!\!\!\!\!{\gamma _{{\rm{Dmax}}}} \!= \!{\gamma _{{\rm{Dteff}}}}\!{\left( {\sum\limits_{l = 1}^L\! {\sum\limits_{m = 1}^{{M_l}}\!\! {{\rho _{{\rm{D}}l}}\!\sqrt {\beta _{lm}^{ - 1}} \left| {{{{\rm{\hat g}}}_{{\rm{D}}lm}}} \right|\!\left| {{{\rm{h}}_{{\rm{D}}lm}}} \right|} } \! +\! {\rho _0}{{{\rm{\hat g}}}_{\rm{0}}}} \!\!\right)^2}.
\end{align}
{\bf{Far-field case:}} In the far-field case, similar to the centralized deployment, we obtain
$d_{lm}^2 \le r_{lm}^{\rm{t}} \approx {d_{1l}}$,  $d_{lm}^2 \le r_{lm}^{\rm{r}} \approx {d_{2l}}$,
$\cos \theta _{lm}^{{\rm{tx}}} \approx 1$ and $\cos \theta _{lm}^{{\rm{rx}}} \approx 1$, which yields
\begin{align}
\label{PL_dis_far}
\beta _{{\rm{D}}l}^{{\rm{farfield}}} \buildrel \Delta \over = \frac{{{\beta _0}{{\left( {{d_{1l}}{d_{2l}}} \right)}^2}}}{{\cos \left( {\theta _l^{\rm{t}}} \right)\cos \left( {\theta _l^{\rm{r}}} \right)}},l = 1, \ldots ,L,
\end{align}
where $\cos \left( {\theta _{lm}^{\rm{t}}} \right)$ and $\cos \left( {\theta _{lm}^{\rm{r}}} \right)$
can be explicitly expressed as $\cos \left( {\theta _{lm}^{\rm{t}}} \right) = {{\left( {{z_{\rm{t}}} - {z_{0l}}} \right)}
\mathord{\left/
 {\vphantom {{\left( {{z_{\rm{t}}} - {z_{0l}}} \right)} {{d_{1l}}}}} \right.
 \kern-\nulldelimiterspace} {{d_{1l}}}}$
 and $\cos \left( {\theta _{lm}^{\rm{r}}} \right) = {{\left( {{z_{\rm{r}}} - {z_{0l}}} \right)} \mathord{\left/
 {\vphantom {{\left( {{z_{\rm{r}}} - {z_{0l}}} \right)} {{d_{2l}}}}} \right.
 \kern-\nulldelimiterspace} {{d_{2l}}}}$, respectively.
Equation \eqref{PL_dis_far} implies that all the $M_l$ elements of the $l$-th RIS experience the same path loss.
Similarly, the received signal in the far-field case can be formulated as
\begin{align}
\!y_{\rm{D}}^{{\rm{farfield}}} &\!= \!\sqrt P\! \left( {\sum\limits_{l = 1}^L \!{{\rho _{{\rm{D}}l}}{{\left( {\sqrt {\beta _{{\rm{D}}l}^{{\rm{farfield}}}} } \right)}^{ - 1}}\!{{{\bf{\hat g}}}_{{\rm{D}}l}}{{\bf{\Phi }}_{{\rm{D}}l}}{{\bf{h}}_{{\rm{D}}l}}}  \!+ \!{\rho _0}{{{\rm{\hat g}}}_0}} \!\!\right)\!s\nonumber\\
& + \underbrace {\sum\limits_{l = 1}^L  \sqrt {P{{\bar \rho }_{{\rm{D}}l}}{{\left( {\beta _{{\rm{D}}l}^{{\rm{farfield}}}} \right)}^{ - 1}}} {{\bm{\omega }}_{{\rm{D}}l}}{{\bf{\Phi }}_{{\rm{D}}l}}{{\bf{h}}_{{\rm{D}}l}}s + {\omega _e}}_{n_{{\rm{Deff}}}^{{\rm{farfield}}}}.
\end{align}
\noindent By co-phasing the signals from all the distributed RIS elements, similar to \eqref{SNR_max_dis},
the optimal received SNR is
\begin{align}
\!\!\!\gamma _{{\rm{Dmax}}}^{{\rm{farfield}}} \!= \!\gamma _{{\rm{Dteff}}}^{{\rm{farfield}}}\!{\left( {\sum\limits_{l = 1}^L \! {\rho _{{\rm{D}}l}}\!{{\left(\!\! {\sqrt {\beta _{{\rm{D}}l}^{{\rm{farfield}}}} } \right)}^{ - 1}}\!\!\sum\limits_{m = 1}^{{M_l}} \! \left| {{{{\rm{\hat g}}}_{{\rm{D}}lm}}} \right|\!\left| {{{\rm{h}}_{{\rm{D}}lm}}} \right| \!+\! {\rho _0}{{{\rm{\hat g}}}_{\rm{0}}}} \!\!\right)^2}.
\end{align}
\section{Performance Analysis}
The objective of this section is to analyze the performance of the centralized and distributed RIS-aided systems based on
$ {\gamma _{{\rm{Cmax}}}}$, $ \gamma _{{\rm{C max }} }^{{\rm{far   field}}}$, $ {\gamma _{{\rm{Dmax}}}}$ and
$ \gamma _{{\rm{Dmax}} }^{{\rm{far   field}}}$.
However, the computation of the exact distribution of these SNRs is mathematically intractable.
To overcome this issue, we introduce approximated expressions of the EC based on the Gamma approximation,
whose tightness is substantiated with the aid of numerical results illustrated in Section IV.
\subsection{Centralized Deployment}
\subsubsection{Gamma Approximation}
Let $\gamma$ be the SNR of interest. The EC normalized by the bandwidth is defined as \cite{ DLiergodic , BAlexergodic , tao2020performance }
\begin{align}
\label{EC_Def}
\!\!\!\bar C \!= \!\mathbb{E}\!\left( {{{\log }_2}\!\left( {1\! +\! \gamma } \right)} \right) \!=\!\! \int_0^\infty \!\! {{{\log }_2}}
\!\left( {1\! + \!\gamma } \right)\!{f_\gamma }\!\left( \gamma  \right)\!d\gamma\;\;{\rm{bit/s/Hz}}.
\end{align}
\begin{coro}\label{coro:Gamma}
Define ${U_m} \!\!\buildrel \Delta \over = \!\!\sqrt {\!\beta _m^{ - 1}}
\left| {{{{\rm{\hat g}}}_{{\rm{C}}m}}} \right|\!\left| {{{\rm{h}}_{{\rm{C}}m}}} \right|$,
$T\! \buildrel \Delta \over = {\rho _{\rm{C}}}\!\!\sum\limits_{m = 1}^M \!\!{{U_m}} $,
$Z{\rm{ }} \buildrel \Delta \over = T + {\rho _0}\left| {{{{\rm{\hat g}}}_0}} \right|$
and $R \buildrel \Delta \over = {Z^2}$. According to \cite[Sec. 2.2.2]{primak2005stochastic},
the probability density function (PDF) of $Z$ can be tightly approximated by a Gamma distribution as follows
\begin{align}
\label{PDFofZ_cen}
{f_Z}\left( z \right) \approx \frac{{b_{Z1}^{{a_{Z1}}}}}{{\Gamma \left( {{a_{Z1}}} \right)}}{z^{{a_{Z1}} - 1}}{e^{ - {b_{Z1}}z}},
\end{align}
where $\Gamma \left( . \right)$
 is the Gamma function as defined in \cite[Eq. (8.310)]{Table}, ${a_{Z1}} \buildrel \Delta \over
 =  \frac{{{{\left( {\mathbb{E}\left( Z \right)} \right)}^2}}}{{\mathrm{Var}\left( Z \right)}}$
, ${b_{Z1}} \buildrel \Delta \over =  \frac{{\mathbb{E}\left( Z \right)}}{{\mathrm{Var}\left( Z \right)}}$
and ${\rm{Var}}\left( Z \right) = \mathbb{E}\left( {{Z^2}} \right) - {\left( {\mathbb{E}\left( Z \right)} \right)^2}$ with
\begin{align}
\label{EZ_cen}
\mathbb{E}\!\left( Z \right) \!\!=\!\! {\rho _{\rm{C}}}\!\!\!\sum\limits_{m = 1}^M\!\! {\sqrt {\!\beta _m^{ - 1}} {\Omega _1}{\Omega _2}}
  \!+\!\! \sqrt {\!\beta _0^{ - 1}} {\rho _0}{\Omega _0},
\end{align}
\begin{align}
\label{EZ^2_cen}
&\mathbb{E}\left( {{Z^2}} \right) = \rho _{\rm{C}}^2\sum\limits_{m = 1}^M {\beta _m^{ - 1} + } \rho _{\rm{C}}^{\rm{2}}
\sum\limits_{m = 1}^M {\sum\limits_{k = 1,k \ne m}^M {\sqrt {\beta _m^{ - 1}\beta _k^{ - 1}}
{{\left( {{\Omega _1}{\Omega _2}} \right)}^2}} } \nonumber\\
& + \beta _0^{ - 1}\rho _0^2 + 2\sqrt {\beta _0^{ - 1}} {\rho _{\rm{C}}}{\rho _0}\sum\limits_{m = 1}^M
 {\sqrt {\beta _m^{ - 1}} {\Omega _0}{\Omega _1}{\Omega _2}}.
\end{align}
In addition, ${\Omega _0}$, ${\Omega _1}$ and ${\Omega _2}$ denote the average values of the Rician variables
$\left| {{{{\rm{\hat h}}}_0}} \right|$,
$\left| {{{\rm{h}}_{{\rm{C}}m}}} \right|$ and $\left|{{{{\rm{\hat g}}}_{{\rm{C}}m}}}\right|$, respectively, which can be expressed as
${\Omega _i} \buildrel \Delta \over = \sqrt {\frac{\pi }{{4\left( {1 + {K_i}} \right)}}} {L_{\frac{1}{2}}}\left( { - {K_i}} \right)$
 $\left( {i \in  \left\{0,1,2\right\}} \right)$ \cite[Eq. (12)]{tao2020performance} with ${L_{\frac{1}{2}}}
\left( . \right)$ being the Laguerre polynomial \cite{abramowitz1972handbook}.
\begin{IEEEproof}
See Appendix A.
\end{IEEEproof}
\end{coro}

By utilizing ${F_Z}\left( z \right) = \int_0^\infty  {{f_Z}\left( z \right)dz} $, we obtain the CDF of $Z$ as
\begin{align}
{F_Z}\left( z \right) = 1 - \frac{{\Gamma \left( {{a_{Z1}},{b_{Z1}}z} \right)}}{{\Gamma \left( {{a_{Z1}}} \right)}}\nonumber.
\end{align}
where $\Gamma \left( {.,.} \right)$ is the incomplete gamma function \cite[Eq. (8.350.2)]{Table}.
Since ${\gamma _{{\rm{Cmax}}}} = {\gamma _{{\rm{Cteff}}}}{Z^2}$, the CDF of ${\gamma _{{\rm{Cmax}}}}$ can be derived by
using the transformation method between two RVs as \cite[Eq. (2.1.49)]{Digital_Commun}
\begin{equation}
\label{CDF_gamma_Cmax}
{F_{{\gamma _{{\rm{Cmax}}}}}}\left( \gamma  \right) = 1 - \frac{1}{{\Gamma \left( {{a_{Z1}}} \right)}}\Gamma \left( {{a_{Z1}},{b_{Z1}}
\sqrt {\frac{\gamma }{{{\gamma _{{\rm{Cteff}}}}}}} } \right).
\end{equation}

\begin{coro}\label{coro: EC_cen_general}
An approximate closed-form expression for the EC of a centralized RIS-aided communication system is given by
\begin{align}
\label{EC_cen_general}
\!\!{{\bar C}_{{\rm{cen}}}} \!\approx \! \frac{{{2^{{a_{Z1}} - 1}}}}{{\sqrt \pi  \Gamma \left( {{a_1}} \right)\ln 2}}G_{3,5}^{5,1}
\!\!\left[ \! {\frac{{b_{Z1}^2}}{{4{\gamma _{{\rm{Cteff}}}}}}\!\!\left|\!\!\!\! {\begin{array}{*{20}{c}}
{\frac{1}{2},1,0}\\
{\frac{{{a_{Z1}}}}{2},\frac{{{a_{Z1}} + 1}}{2},0,\frac{1}{2},0}
\end{array}} \right.} \!\!\!\right],
\end{align}
where  $G\left( . \right)$ is the Meijer's $G$-function \cite[Eq. (9.301)]{Table}.
In addition, ${\gamma _{{\rm{Cteff}}}} \buildrel \Delta \over = \frac{P}{{\sigma _{{\rm{Ceff}}}^2}}$
with ${\sigma _{{\rm{Ceff}}}^2} $ defined as
\begin{align}
\label{sigma_Ceff}
\!\sigma _{{\rm{Ceff}}}^2\buildrel \Delta \over =\mathbb{E}\!\left( {{{\left| {{n_{{\rm{Ceff}}}}} \right|}^2}}\! \right) &= P\bar \rho _{\rm{C}}^2\!\left( {1 - \Omega _2^2} \right)\sum\limits_{m = 1}^M {\beta _m^{ - 1}} \nonumber \\
&+P\bar \rho _{\rm{0}}^2\left( {1- \Omega _0^2} \right)\beta _0^{ - 1} + \sigma _0^2.
\end{align}
\begin{IEEEproof}
See Appendix B.
\end{IEEEproof}
\end{coro}
\noindent By using \eqref{sigma_Ceff} and ${\gamma _{{\rm{Cteff}}}} \buildrel \Delta \over = {P \mathord{\left/
 {\vphantom {P {E\left( {{{\left| {{n_{{\rm{Ceff}}}}} \right|}^2}} \right)}}} \right.
 \kern-\nulldelimiterspace} {\mathbb{E}\left( {{{\left| {{n_{{\rm{Ceff}}}}} \right|}^2}} \right)}}$, we obtain \eqref{29}, from which we conclude that ${\gamma _{{\rm{Cteff}}}}$ tends to be constant when $P$ tends to infinity.
 \begin{align}
 \label{29}
 \!\!\!\!\!\!\!{\gamma _{{\rm{Cteff}}}} \!= \!{P\!\!\! \mathord{\left/
 {\vphantom {P\!\! {\left( \!\!{P\!\!\bar \rho _{\rm{C}}^2\!\left( {1 \!-\! \Omega _2^2} \right)\!\!\sum\limits_{m = 1}^M \! {\beta _m^{ - 1}}  \!+\! P\bar \rho _{\rm{0}}^2\!\left(\! {1\! -\! \Omega _0^2} \right)\!\beta _0^{ - 1}\! +\! \sigma _0^2} \right)}}} \right.
 \kern-\nulldelimiterspace} \!\!\!{\left(\! \!{P\bar \rho _{\rm{C}}^2\!\left( \!{1 \!-\! \Omega _2^2} \right)\!\!\sum\limits_{m = 1}^M\!\! {\beta _m^{ - 1}}\!  +\! P\bar \rho _{\rm{0}}^2\!\left( \!{1\! -\! \Omega _0^2} \right)\!\beta _0^{ - 1}\! +\! \sigma _0^2} \!\!\right)}}.
 \end{align}
{\bf{Far-field case:}} In the far-field regime, from (9), we have ${\beta _m} = {\beta ^{{\rm{far  field}}}}$ for $m = 1, \ldots ,M$.
With the aid of analytical steps similar to those of Corollary 1 and Corollary 2, the EC can be formulated as
\begin{align}
\label{EC_cen_far}
\!\!\!\!\!\!\bar C_{{\rm{cen}}}^{{\rm{far  field}}} \! \approx \!\! \frac{{{2^{{a_2} - 1}}}}
{{\sqrt \pi  \Gamma\! \left(\! {{a_{Z2}}}\! \right)\!\ln\! 2}}G_{3,5}^{5,1}
\!\!\left[\! {\frac{{b_{Z2}^2}}{{4\gamma _{{\rm{Cteff}}}^{{\rm{far  field}}}}}\!\!\left|\!\!\!\! {\begin{array}{*{20}{c}}
{\frac{1}{2},1,0}\\
{\frac{{{a_{Z2}}}}{2},\!\frac{{{a_{Z2}} + 1}}{2},0,\frac{1}{2},0}
\end{array}} \right.} \!\!\!\!\!\right]\!,
\end{align}
where ${a_{Z2}} \buildrel \Delta \over = \frac{{\mathbb{E}\left( {{{\left( {{Z^{{\rm{farfield}}}}} \right)}^2}} \right)}}
{{{\rm{Var}}\left( {{Z^{{\rm{farfield}}}}} \right)}}$,
${b_{Z2}} \buildrel \Delta \over = \frac{{\mathbb{E}\left( {{Z^{{\rm{farfield}}}} } \right)}}
{{{\rm{Var}}\left( {{Z^{{\rm{farfield}}}}} \right)}}$,
$\gamma _{{\rm{Cteff}}}^{{\rm{far field}}} \buildrel \Delta \over = \frac{P}
{{{{\left( {\sigma _{{\rm{Ceff}}}^{{\rm{farfield}}}} \right)}^2}}}$ and

\begin{align}
&\mathbb{E}\left( {{Z^{{\rm{farfield}}}}} \right) = \frac{{M{\rho _{\rm{C}}}{\Omega _1}
{\Omega _2}}}{{\sqrt {{\beta ^{{\rm{far  field}}}}} }}
+ \frac{{{\rho _0}{\Omega _0}}}{{\sqrt {{\beta _0}} }},
\end{align}
\begin{align}
\mathbb{E}\left( {{{\left( {{Z^{{\rm{farfield}}}}} \right)}^2}} \right) &= \frac{{M\rho _{\rm{C}}^2}}{{{\beta ^{{\rm{farfield}}}}}}
+ \frac{{2M{\rho _{\rm{C}}}{\rho _0}{\Omega _0}{\Omega _1}{\Omega _2}}}{{\sqrt {{\beta _0}{\beta ^{{\rm{farfield}}}}} }}\nonumber\\
 &+ \frac{{M\left( {M - 1} \right)\rho _{\rm{C}}^2{{\left( {{\Omega _1}{\Omega _2}} \right)}^2}}}{{{\beta ^{{\rm{farfield}}}}}}
+ \frac{{\rho _0^2}}{{{\beta _0}}},
\end{align}
\begin{align}
{\left( {\sigma _{{\rm{Ceff}}}^{{\rm{farfield}}}} \right)^2} = \frac{{PM\bar \rho _{\rm{C}}^2\left( {1 - \Omega _2^2} \right)}}{{{\beta ^{{\rm{farfield}}}}}} + \frac{{P\bar \rho _{\rm{0}}^2\left( {1 - \Omega _0^2} \right)}}{{{\beta _0}}} + \sigma _0^2.\nonumber
\end{align}
If $\rho_\mathrm{C}=1$ and ${{\rm{h}}_0}=0$ (i.e., no outdated CSI and no direct link),
the EC reduces to that obtained in \cite{salhab2021accurate}, as expected.

\subsubsection{Bounds for the EC}
Although \eqref{EC_cen_general} can be utilized to efficiently evaluate the EC,
it is difficult to explicitly analyze the impact of the system and channel parameters on the achievable performance.
Thus, to gain useful design insights, we provide tight upper and lower bounds for the EC by using Jensen's inequality.
In particular, the following upper and lower bounds are considered
\begin{align}
\label{Jensen}
\!\!{\bar C^{{\rm{lb}}}} \!\buildrel \Delta \over =\! {\log _2}\!\!\left(\! {1\!\! +\! \!{{\left( {\mathbb{E}\!
\left( {{1 \mathord{\left/
 {\vphantom {1 \gamma }} \right.
 \kern-\nulldelimiterspace} \gamma }} \right)} \right)}^{ - 1}}} \!\right) \!\le\! C \!\!\le\!\! {\bar C^{{\rm{ub}}}}
 \buildrel \Delta \over =
 {\log _2}\!\left(\! {1 \!+\! \mathbb{E}\!\left( \gamma  \right)} \!\right).
\end{align}
\begin{coro}\label{coro:Cen_EC_bound}
The EC of a centralized RIS-aided communication systems is upper bounded by
${\bar C^{{\rm{ub}}}} = {\log _2}\left( {1 + {\mathbb{E}}\left( {{\gamma _{{\rm{Cmax}}}}} \right)} \right)$, where
\begin{align}
\label{EC_cen_ub}
\!\!\!\!\!\mathbb{E}\!\left( {{\gamma _{{\rm{Cmax}}}}} \right) \!\!&=\!\! {\gamma _{{\rm{Cteff}}}}\!\!\left( \!\!{\rho _{\rm{C}}^2
\!\!\sum\limits_{m = 1}^M\!\! {\beta _m^{ - 1}}  \!+\! \rho _{\rm{C}}^2\Omega _1^2\Omega _2^2\!\sum\limits_{m = 1}^M \!
{\sum\limits_{k = 1,k \ne m}^M\!\!\! {\sqrt {\!\beta _m^{ - 1}\!\beta _k^{ - 1}} } } } \right.\nonumber\\
&\left. { \!+\! 2{\rho _{\rm{C}}}{\rho _0}{\Omega _0}{\Omega _1}{\Omega _2}\sqrt {\!\beta _0^{ - 1}}\!
\sum\limits_{m = 1}^M\!\! {\sqrt {\beta _m^{ - 1}} } + \rho _0^2{\beta _0^{-1}} } \right).
\end{align}
Also, the lower bound can be approximated as
\begin{align}
\label{lowerbound}
\!\!{{\bar C}^{{\rm{lb}}}}\! \approx\! {\log _2}\!\left\{ {1\! +\! {{\left( {\frac{1}{{\mathbb{E}\left( {{\gamma _{{\rm{Cmax}}}}} \right)}}
\!+ \!\frac{{{\rm{Var}}\left( {{\gamma _{{\rm{Cmax}}}}} \right)}}
{{{{\left( {\mathbb{E}\left( {{\gamma _{{\rm{Cmax}}}}} \right)} \right)}^3}}}} \right)}^{ - 1}}} \right\},
\end{align}
where
\begin{align}
\label{Variance}
\!\!{\rm{Var}}\left( {{\gamma _{{\rm{Cmax}}}}} \right)\! \approx \!\frac{{\gamma _{{\rm{Cteff}}}^2}}{{b_{Z1}^4}}
\!\left[ {\frac{{\Gamma \left( {{a_{Z1}} \!+ \!4} \right)}}{{\Gamma \left( {{a_{Z1}}} \right)}} \!-\! \frac{{{\Gamma ^2}
\left( {{a_{Z1}}\! +\! 2} \right)}}{{{\Gamma ^2}\!\left( {{a_{Z1}}} \right)}}} \right].
\end{align}
\begin{IEEEproof}
See Appendix C.
\end{IEEEproof}
\end{coro}
\begin{remark}
By direct inspection of (32) and (33), we observe that the EC increases when ${\gamma _{{\rm{Cteff}}}}$,
${{\rho _{\rm{0}}}}$ and/or ${{\rho _{\rm{C}}}}$ increase.
Since ${L_{\frac{1}{2}}}\left( { - K_i} \right)$ is a monotonically increasing function of $K_i$, in addition, ${\Omega _i}$ increases with $K_i$, which suggests that the EC is enhanced in the presence of a strong LoS component. If ${K_i} \to \infty $, i.e., only the LoS components exist in the considered Rician fading channel model, however, the EC of the RIS-assisted system tends to a constant, since, by definition, ${\Omega _i} \to 1$ when $K_i$ is sufficiently large.
If $K_1=0$ or $K_2=0$ and $K_0=0$, furthermore, we obtain ${\Omega _i} = {{\sqrt \pi } \mathord{\left/ {\vphantom {{\sqrt \pi  } 2}} \right. \kern-\nulldelimiterspace} 2}$. We observe that when the cascaded and direct channels are subject to Rayleigh fading, the EC increases with $M$, $\rho_0$ and $\rho_\mathrm{C}$. If $P \to \infty $, also, $ {\gamma _{{\rm{Cteff}}}}$ tends to a constant. This reveals that the EC does not increase without bound with the transmit power.
\end{remark}


{\bf{Far-field case:}} In the far-field case, we have
\begin{align}
\label{SNR_max_far}
\mathbb{E}\!\!\left(\! {\gamma _{{\rm{Cmax}}}^{{\rm{farfield}}}} \!\right)\!& =\! \gamma _{{\rm{Cteff}}}^{{\rm{farfield}}}
\!\!\left( {{{M\!\rho _{\rm{C}}^2}\! \mathord{\left/
 {\vphantom {{M\rho _{\rm{C}}^2} {{\beta ^{{\rm{farfield}}}}}}} \right.
 \kern-\nulldelimiterspace} \!{{\beta ^{{\rm{farfield}}}}}} \!+\! {{M\!\!\left( \!{M \!-\! 1}\! \right)
 \!\rho _{\rm{C}}^2\Omega _1^2\Omega _2^2} \!\mathord{\left/
 {\vphantom {{M\left( {M - 1} \right)\rho _{\rm{C}}^2\Omega _1^2\Omega _2^2}\!\! {{\beta ^{{\rm{far - field}}}}}}} \right.
 \kern-\nulldelimiterspace}\! {{\beta ^{{\rm{farfield}}}}}}} \right.\nonumber\\
&\left. { + {{2M{\rho _{\rm{C}}}{\rho _0}{\Omega _0}{\Omega _1}{\Omega _2}} \!\mathord{\left/
 {\vphantom {{2M{\rho _{\rm{C}}}{\rho _0}{\Omega _0}{\Omega _1}{\Omega _2}} {\sqrt {{\beta _0}{\beta ^{{\rm{far - field}}}}} }}} \right.
 \kern-\nulldelimiterspace}\!\! {\sqrt {{\!\beta _0}{\beta ^{{\rm{farfield}}}}} }} \!+ \rho _0^2\beta _0^{ - 1}} \right),
\end{align}
\begin{align}
\label{variance_far_cen}
\!\!\!\!\!{\rm{Var}}\!\left( {\gamma _{{\rm{Cmax}}}^{{\rm{far field}}}} \right) \!
\approx \!\frac{{{{\left( {\gamma _{{\rm{Cteff}}}^{{\rm{far field}}}}
 \right)}^2}}}{{b_{Z2}^4}}\!\!\left[ {\frac{{\Gamma \!\left( {{a_{Z2}} \!+\! 4} \right)}}{{\Gamma \left( {{a_{Z2}}} \right)}} \!
 -\! \frac{{{\Gamma ^2}\!\left( {{a_{Z2}}\! +\! 2} \right)}}{{{\Gamma ^2}\left( {{a_{Z2}}} \right)}}} \right].
\end{align}
Substituting \eqref{SNR_max_far} into \eqref{Jensen} and substituting \eqref{variance_far_cen} into \eqref{lowerbound},
we obtain the upper and lower bounds for the EC in the far-field case, respectively.
By setting $K_0=0$ and $\rho_0=\rho_\mathrm{C}=1$ as a special case,
we retrieve the upper bound for the EC in the absence of outdated CSI over a Rayleigh fading channel \cite{tao2020performance}.
Although \eqref{SNR_max_far} indicates that the EC increases with the number of elements of the RIS,
it needs to be noted that the far-field assumption may no longer hold if $M$ is very large.
In this case, it is necessary to use \eqref{EC_cen_ub} to accurately estimating the EC.

\subsection{Distributed Deployment}
\subsubsection{Gamma approximation}
\begin{coro}\label{coro:gamma_Dis}
Define ${X_{lm}} \buildrel \Delta \over = \sqrt {\beta _{lm}^{ - 1}} \left| {{\mathrm{{\hat g}}_{{\rm{D}}lm}}} \right|
\left| {{\mathrm{h}_{{\rm{D}}lm}}} \right|$,
 $Y \buildrel \Delta \over = \sum\limits_{l = 1}^L {\sum\limits_{m = 1}^{{M_l}} {{\rho _{{\rm{D}}l}}{X_{lm}}} } $
 and $H \buildrel \Delta \over = Y +  {\rho _0}\left| {{{{\rm{\hat g}}}_0}} \right|$,
 then the PDF of $H$ can be tightly approximated by a Gamma distribution as follows
 \begin{align}
 \label{PDFofH}
 {f_H}\left( h \right) \approx \frac{{b_{H1}^{{a_{H1}}}}}{{\Gamma \left( {{a_{H1}}} \right)}}{h^{{a_{H1}} - 1}}{e^{ - {b_{H1}}h}},
 \end{align}
where ${a_{H1}} \buildrel \Delta \over = \frac{{{{\left( {\mathbb{E}\left( H \right)} \right)}^2}}}{{{\rm{Var}}\left( H \right)}}$,
${b_{H1}} \buildrel \Delta \over = \frac{{\mathbb{E}\left( H \right)}}{{{\rm{Var}}\left( H \right)}}$
and ${\rm{Var}}\left( H \right) = \mathbb{E}\left( {{H^2}} \right) - {\left( {\mathbb{E}\left( H \right)} \right)^2}$ with
\begin{align}
\label{EH}
\!\!\!\!&\mathbb{E}\!\left( H \right) \!=\!\! \sum\limits_{l = 1}^L \!{\sum\limits_{m = 1}^{{M_l}}\!\! {\sqrt {\!\beta _{lm}^{ - 1}}
{\rho _{{\rm{D}}l}}{\Omega _{1l}}{\Omega _{2l}}} } \! +\! \sqrt {\!\beta _0^{ - 1}} {\rho _0}{\Omega _0}.
\end{align}
\begin{align}
\label{C9}
\mathbb{E}\!\left( {{H^2}} \right) \!\!&= \sum\limits_{l = 1}^L {\sum\limits_{m = 1}^{{M_l}} {\frac{{\rho _{{\rm{D}}l}^2}}{{{\beta _{lm}}}}} }
+ \sum\limits_{l = 1}^L {\sum\limits_{m = 1}^{{M_l}} {\sum\limits_{k = 1,k \ne m}^{{M_l}}
{\frac{{\rho _{{\rm{D}}l}^2\Omega _{1l}^2\Omega _{2l}^2}}{{\sqrt {{\beta _{lm}}{\beta _{lk}}} }}} } }\nonumber \\
&+\sum\limits_{l = 1}^L {\sum\limits_{m = 1}^{{M_l}} {\left( {\frac{{{\rho _{{\rm{D}}l}}{\Omega _{1l}}
{\Omega _{2l}}}}{{\sqrt {{\beta _{lm}}} }}\sum\limits_{j = 1,j \ne l}^L {\sum\limits_{k = 1}^{{M_j}}
{\frac{{{\rho _{{\rm{D}}j}}{\Omega _{1j}}{\Omega _{2j}}}}{{\sqrt {{\beta _{jk}}} }}} } } \right)} }\nonumber \\
 &+ \frac{{2{\rho _0}}}{{\sqrt {{\beta _0}} }}\sum\limits_{l = 1}^L {\sum\limits_{m = 1}^{{M_l}}
{\frac{{{\rho _{{\rm{D}}l}}{\Omega _{1l}}{\Omega _{2l}}}}{{\sqrt {{\beta _{lm}}} }} + \frac{{\rho _0^2}}{{{\beta _0}}}} },
\end{align}
where ${\Omega _{1l}}$ and ${\Omega _{2l}}$ denote the average values of the Rician variables
$\left|{{{\rm{h}}_{{\rm{D}}lm}}}\right|$ and $\left|{{{{\rm{\hat g}}}_{{\rm{D}}lm}}}\right|$, respectively, which can be expressed as
${\Omega _{il}} \buildrel \Delta \over = \sqrt {\frac{\pi }{{4\left( {1 + {K_{il}}} \right)}}} {L_{\frac{1}{2}}}\left( { - {K_{il}}} \right)$
 $\left( {i \in \left\{1,2\right\}} \right)$.
 \begin{IEEEproof}
See Appendix D.
\end{IEEEproof}
\end{coro}
By exploiting a similar methodology as for the derivation of \eqref{CDF_gamma_Cmax}, we arrive at
\begin{equation}
\label{cdf_gamma_Dmax}
{F_{{\gamma _{{\rm{Dmax}}}}}}\left( \gamma  \right) = 1 - \frac{1}{{\Gamma \left( {{a_{H1}}} \right)}}
\Gamma \left( {{a_{H1}},{b_{H1}}\sqrt {\frac{\gamma }{{{\gamma _{{\rm{Dteff}}}}}}} } \right).
\end{equation}

\begin{coro}
\label{coro:EC_dis}
An approximated closed-form expression for the EC of a distributed RIS-aided communication system is given by
\begin{align}
\label{EC_dis_near_gamma}
\!\!\!{{\bar C}_{{\rm{dis}}}} \!\approx \! \frac{{{2^{{a_{H1}} - 1}}}}{{\sqrt \pi  \Gamma \!\left(\! {{a_{H1}}} \!\right)\ln 2}}\!
G_{3,5}^{5,1}\!\!\left[\! {\frac{{b_{H1}^2}}{{4{\gamma _{{\rm{Dteff}}}}}}\!\left|\!\!\! {\begin{array}{*{20}{c}}
{\frac{1}{2},1,0}\\
{\frac{{{a_{H1}}}}{2},\frac{{{a_{H1}} + 1}}{2},0,\frac{1}{2},0}
\end{array}} \right.}\!\!\!\! \right],
\end{align}
where
${\gamma _{{\rm{Dteff}}}} \buildrel \Delta \over = {P \mathord{\left/
 {\vphantom {P {\sigma _{{\rm{Deff}}}^2}}} \right.
 \kern-\nulldelimiterspace} {\sigma _{{\rm{Deff}}}^2}}$ and
\begin{align}
\label{noise_variance_dis}
\!\!\sigma _{{\rm{Deff}}}^2\! \buildrel \Delta \over = \!P\!\sum\limits_{l = 1}^L\!  \sum\limits_{m = 1}^{{M_l}} \! \frac{{\bar \rho _{{\rm{D}}l}^2\!\left( {1 \!-\! \Omega _{2l}^2} \right)}}{{{\beta _{lm}}}} \!+ \!\frac{{P\bar \rho _{\rm{0}}^2\!\left( {1\! - \!\Omega _0^2} \right)}}{{{\beta _0}}} \!+ \!\sigma _0^2.
\end{align}
\begin{IEEEproof}
See  Appendix E.
\end{IEEEproof}
\end{coro}
By setting $L=1$, \eqref{EC_dis_near_gamma} reduces to the EC of the centralized RIS deployment in \eqref{EC_cen_general}

{\bf{Far-field case:}} By setting ${\beta _{lm}} = \beta _{{\rm{D}}l}^{{\rm{farfield}}}$,
we can obtain the EC for the distributed deployment in the far-field regime
\begin{align}
\!\!\!\!\!\!\!\bar C_{{\rm{dis}}}^{{\rm{farfield}}} \!=\!\! \frac{{{2^{{a_{H2}} - 1}}}}{{\sqrt \pi  \Gamma\! \left(\! {{a_{H2}}} \!\right)\!\ln \! 2}}\!
G_{3,5}^{5,1}\!\!\left[\! {\frac{{b_{H2}^2}}{{4\gamma _{{\rm{Dteff}}}^{{\rm{farfield}}}}}\!\!\left|\!\!\! {\begin{array}{*{20}{c}}
{\frac{1}{2},1,0}\\
{\frac{{{a_{H2}}}}{2}\!,\!\frac{{{a_{H2}} + 1}}{2}\!,\!0,\frac{1}{2},\!0}
\end{array}} \right.} \!\!\!\!\right]\!,
\end{align}
where ${a_{H2}} \buildrel \Delta \over = \frac{{{{\left( {\mathbb{E}\left( {{H^{{\rm{farfield}}}}} \right)} \right)}^2}}}
{{{\rm{Var}}\left({{H^{{\rm{farfield}}}}} \right)}}$ and ${b_{H2}} \buildrel \Delta \over = \frac{{E\left( {{H^{{\rm{farfield}}}}} \right)}}
{{{\rm{Var}}\left( {{H^{{\rm{farfield}}}}} \right)}}$. Moreover, we have
\begin{align}
\mathbb{E}\left( {{H^{{\rm{farfield}}}}} \right) = \sum\limits_{l = 1}^L
{\sum\limits_{m = 1}^{{M_l}} {\frac{{{\rho _{{\rm{D}}l}}
{\Omega _{1l}}{\Omega _{2l}}}}{{\sqrt {\beta _{{\rm{D}}l}^{{\rm{farfield}}}} }}} }
+ \frac{{{\rho _0}{\Omega _0}}}{{\sqrt {{\beta _0}} }},
\end{align}
\begin{align}
&\mathbb{E}\!\!\left(\!\! {{{\left( {{H^{{\rm{farfield}}}}} \right)}^2}}\! \right)\!\! =\!\! \frac{{\rho _0^2}}{{\sqrt {{\!\beta _0}} }}
\!\!+ \!\!\!\sum\limits_{l = 1}^L \!\!{\frac{{{M_l}\rho _{{\rm{D}}l}^2}}{{\sqrt {\!\beta _{{\rm{D}}l}^{{\rm{farfield}}}} }}}  \!\!
+\!\! \sum\limits_{l = 1}^L \!{\frac{{{M_l}\!\left( {{M_l} \!-\! 1} \right)\rho _{\mathrm{D}l}^2\Omega _{1l}^2\Omega _{2l}^2}}
{{\beta _{{\rm{D}}l}^{{\rm{farfield}}}}}}\nonumber \\
& \!\!+\!\! 2\!\!\sum\limits_{l = 1}^L \!\!{\frac{{{M_l}{\rho _{{\rm{D}}l}}{\rho _0}{\Omega _0}{\Omega _{1l}}{\Omega _{2l}}}}
{{\sqrt {{\beta _0}\beta _{{\rm{D}}l}^{{\rm{farfield}}}} }}} \!\!+\!\!\!\sum\limits_{l = 1}^L\!\!
{\left( \!\!\!{\frac{{{M_l}{\rho _{{\rm{D}}l}}{\Omega _{1l}}{\Omega _{2l}}}}{{\sqrt {\!\beta _{{\rm{D}}l}^{{\rm{farfield}}}} }}\!\!\!\!\!
\sum\limits_{j = 1,j \ne l}^L \!\!\!\!\!{\frac{{{M_j}{\rho _{{\rm{D}}j}}{\Omega _{1j}}
{\Omega _{2j}}}}{{\sqrt {\!\beta _{{\rm{D}}j}^{{\rm{farfield}}}} }}} }\!\!\! \right)}\!.\nonumber
\end{align}

\subsubsection{Bounds for the EC}
\begin{coro}\label{coro:Dis_EC_bound}
The EC of a decentralized RIS-aided communication systems is upper bounded by
\begin{align}
\label{EC_dis_ub}
\!\!\!\!\!\mathbb{E}\!\left( {{\gamma _{{\rm{Dmax}}}}} \right)\! &= \!\gamma _{{\rm{Deff}}}\!\!\left( {\frac{{\rho _0^2}}{{{\beta _0}}}
\!+ \!\sum\limits_{l = 1}^L {\sum\limits_{m = 1}^{{M_l}}\! {\frac{{\rho _{Dl}^2}}{{{\beta _{lm}}}}} }
\!+\! 2\!\sum\limits_{l = 1}^L \!{\sum\limits_{m = 1}^{{M_l}}\! {\frac{{{\rho _{{\rm{D}}l}}{\rho _0}{\Omega _0}
{\Omega _{1l}}{\Omega _{2l}}}}{{\sqrt {{\beta _0}{\beta _{lm}}} }}} } } \right.\nonumber\\
& + \sum\limits_{l = 1}^L {\sum\limits_{m = 1}^{{M_l}}
{\sum\limits_{k = 1,k \ne m}^{{M_l}} {\frac{{\rho _{{\rm{D}}l}^2\Omega _{1l}^2\Omega _{2l}^2}}{{\sqrt {{\beta _{lm}}{\beta _{lk}}} }}} } }\nonumber\\
&\left. { \!+ \!\!\sum\limits_{l = 1}^L  \!\sum\limits_{m = 1}^{{M_l}}\!\!  \left( \!\!{\frac{{{\rho _{{\rm{D}}l}}{\Omega _{1l}}{\Omega _{2l}}}}
{{\sqrt {{\beta _{lm}}} }}\!\!\!\!\sum\limits_{j = 1,j \ne l}^L \sum\limits_{k = 1}^{{M_j}}\!\!
\!\frac{{{\rho _{{\rm{D}}j}}{\Omega _{1j}}{\Omega _{2j}}}}{{\sqrt {{\beta _{jk}}} }}}\!\! \right)}\!\!\! \right)\!\!.
\end{align}
Also, the lower bound can be approximated as
\begin{align}
{\rm{Var}}\left( {{\gamma _{{\rm{D max}} }}} \right) \!=\! \frac{{\gamma _{{\rm{Dteff}}}^{\rm{2}}}}{{b_{H1}^4}}\!\!
\left[ {\frac{{\Gamma \left( {{a_{H1}} + 4} \right)}}{{\Gamma \left( {{a_{H1}}} \right)}} \!
-\! \frac{{{\Gamma ^2}\left( {{a_{H1}} + 2} \right)}}{{{\Gamma ^2}\left( {{a_{H1}}} \right)}}} \right].
\end{align}
\begin{IEEEproof}
See Appendix F.
\end{IEEEproof}
\end{coro}
{\bf{Far-field case:}} In the far-field case, the EC can be approximated as follows
\begin{small}
\begin{align}
\label{SNR_max_far_dis}
\!\mathbb{E}\left( {\gamma _{{\rm{Dmax}}}^{{\rm{farfield}}}} \right)\! &=\! \gamma _{{\rm{Deff}}}^{{\rm{farfield}}}\!\left( {\frac{{\rho _0^2}}{{{\beta _0}}}\! + \!\!\sum\limits_{l = 1}^L  \!\frac{{{M_l}\rho _{{\rm{D}}l}^2}}{{\beta _{{\rm{D}}l}^{{\rm{farfield}}}}}} \right. \!+\! 2\sum\limits_{l = 1}^L \!{\frac{{{M_l}{\rho _{{\rm{D}}l}}{\rho _0}{\Omega _0}{\Omega _{1l}}{\Omega _{2l}}}}{{\sqrt {{\beta _0}\!\beta _{{\rm{D}}l}^{{\rm{farfield}}}} }}} \nonumber\\
& + \sum\limits_{l = 1}^L {\frac{{{M_l}\left( {{M_l} - 1} \right)\rho _{{\rm{D}}l}^2\Omega _{1l}^2\Omega _{2l}^2}}{{\beta _{{\rm{D}}l}^{{\rm{farfield}}}}}} \nonumber\\
&\left. { \!+\! \sum\limits_{l = 1}^L \!\!{\left(\! {\frac{{{M_l}{\rho _{{\rm{D}}l}}{\Omega _{1l}}{\Omega _{2l}}}}{{\sqrt {\beta _{{\rm{D}}l}^{{\rm{farfield}}}} }}\!\!\!\sum\limits_{j = 1,j \ne l}^L {\frac{{{M_j}{\rho _{{\rm{D}}j}}{\Omega _{1j}}{\Omega _{2j}}}}{{\sqrt {\beta _{{\rm{D}}j}^{{\rm{farfield}}}} }}} } \right)} } \!\!\right),
\end{align}
\begin{align}
\label{variance_far_dis}
{\rm{Var}}\left( {\gamma _{{\rm{Dmax}}}^{{\rm{far field}}}} \right) \!\approx \!\frac{{{{\left( {\gamma _{{\rm{Dteff}}}^{{\rm{far field}}}}
 \right)}^2}}}{{b_{H2}^4}}\!\!\left[ {\frac{{\Gamma \left( {{a_{H2}} \!+\! 4} \right)}}{{\Gamma \left( {{a_{H2}}} \right)}} \!
 -\! \frac{{{\Gamma ^2}\left( {{a_{H2}}\! +\! 2} \right)}}{{{\Gamma ^2}\left( {{a_{H2}}} \right)}}} \right].
\end{align}
\end{small}
From \eqref{SNR_max_far_dis}, we can infer similar performance trends as for the centralized deployment.

\section{Numerical Results}
In this section, we compare the analytical results against Monte Carlo simulations.
The simulation parameters are provided in Table I. In addition,
the path loss of the direct link is modeled as
$\beta _0^{ - 1}\left[ {{\rm{dB}}} \right] = \eta  - 10\xi {\log _{10}}\left( {{d_0}} \right)$,
where $\eta  =  - 30{\rm{dB}}$ is a reference path loss, $\xi  = 3.5$ is the path loss exponent,
and ${{d_0}}$ is the distance from the BS to the user.
In Fig. 2, the accuracy of the Gamma approximation as well as the tightness of the upper and lower bounds for the EC are examined. In the centralized deployment, the RIS is located in $\left( { - 49.5,0,9.5} \right)$. The size of all the RIS elements is ${\lambda  \mathord{\left/
 {\vphantom {\lambda  8}} \right.
 \kern-\nulldelimiterspace} 8}{{ \times \lambda } \mathord{\left/
 {\vphantom {{ \times \lambda } 8}} \right.
 \kern-\nulldelimiterspace} 8}$.
It can be observed from Fig. 2 that the Gamma approximation provides an almost perfect match with the simulated results.
Furthermore, it can be seen that the performance gap between the upper and lower bounds and Monte Carlo simulations diminishes as the number of reflecting elements $M$ increases, which confirms the accuracy of the bounds.
Moreover, we observe that the EC increases as the Rician-$K$ factor of the direct link increases. In addition, the gap between the Monte Carlo results and the upper and lower bounds decreases with the increase of the Rician $K$-factor.
Also, it can be seen that the EC first improves with an increase of the transmit power $P$. When $P$ is sufficiently large, however, the capacity tends to a limit, which can be explained by the fact that the equivalent SNR tends to be constant as $P$ increases. The results validate the correctness of {\it{Remark 1}}.
\begin{table}[t]
  \label{TABLE_I}
  \centering
  \caption{Simulation parameters}
  \label{table}
  \begin{tabular}{|c|c|}
  \hline
  Parameter & Value \\
  \hline
  Location of the BS $\left( {{x_{{\rm{t}}}},{y_{{\rm{t}}}},{z_{{\rm{t}}}}} \right)$ & $\left( { - 50,0,10} \right)$ m  \\
  \hline
  Location of the user $\left( {{x_{{\rm{r}}}},{y_{{\rm{r}}}},{z_{{\rm{r}}}}} \right)$ & $\left( { 50,0,10} \right)$ m  \\
  \hline
  Carrier frequency & ${f_c} = 5$ GHz\\
  \hline
  Noise power & $\sigma _0^2 =  - 120$ dBm\\
  \hline
  Size of RIS elements & ${d_x} = {d_y} \in \left[ {{\lambda  \mathord{\left/
 {\vphantom {\lambda  {10}}} \right.
 \kern-\nulldelimiterspace} {10}},{\lambda  \mathord{\left/
 {\vphantom {\lambda  2}} \right.
 \kern-\nulldelimiterspace} 2}} \right]$\\
 \hline
 Antenna gains at the BS and user &${G_t} = 20$ dB, $ {G_r} = 0$ dB\\
  \hline
\end{tabular}
\end{table}

In Fig. 3, we analyze the EC as a function of the correlation coefficient $\rho$ and the size $d_x \times d_y$ of the reflecting elements of RIS. For the centralized deployment, the RIS is located in $\left( { - 49.5,0,9.5} \right)$ with $M = {M_x} \times {M_y} = 24 \times 24 = 576$ elements.
For the distributed deployment, on the other hand, we consider two RISs located in $\left( { - 49,0,9.5} \right)$ and $\left( {  49,0,9.5} \right)$, respectively. In addition, the two RISs are equipped with ${M_{xi}} \times {M_{yi}} = 16 \times 18 = 288$ $\left( {i \in \left\{1,2\right\}} \right)$ elements, thus the total number of RIS elements is the same as for the centralized deployment. We assume ${\rho _1} = {\rho _2} = \rho $.
It can be observed from Fig. 3 that the accuracy of the bounds and the Gamma approximation is demonstrated.
Furthermore, the EC decreases as $\rho$ decreases (i.e., the CSI becomes more outdated) for both the centralized and distributed deployments.
Moreover, we find that increasing the size of the reflecting elements can significantly improve the system performance.

In Fig. 4, we present the EC in the near- and far-field regions for the centralized RIS deployment by using
the general expression in \eqref{EC_cen_general} and the far-field expression in \eqref{EC_cen_far}.
The boundary between the near-field and the far-field region is calculated as
${D_{{\rm{boundary}}}} \!=\! {{2\left( {{{\left( {{M_x}{d_x}} \right)}^2} + {{\left( {{M_y}{d_y}} \right)}^2}} \right)} \mathord{\left/
 {\vphantom {{2\left( {{{\left( {{M_x}{d_x}} \right)}^2} + {{\left( {{M_y}{d_y}} \right)}^2}} \right)} \lambda }} \right.
 \kern-\nulldelimiterspace} \lambda }$. For example, ${D_{{\rm{boundary}}}}$ can be calculated as $6$ m, $4.7$ m, and $3.75$ m for $M_x=M_y=40$, $M_x=30, M_y=40$, and $M_x=20, M_y=40$ if $d_x=d_y={{\lambda  \mathord{\left/
 {\vphantom {\lambda  8}} \right.
 \kern-\nulldelimiterspace} 8}}$, respectively.
As observed from the figure, in the near-field region, there is a performance gap between the two formulas,
and the gap increases with the increase of the total number of RIS elements.
This indicates that it is inaccurate to use the far-field formula to analyze the EC if the BS or the user is in the near-field of the RIS.
Furthermore, we observe that the far-field formula gradually approaches the general formula as the distance between the
BS and the RIS increases: when ${d_1} > {D_{{\rm{boundary}}}}$, the two curves almost coincide.

Figure 5 illustrates the impact of the size of the unit cells (${d_x} \times {d_y}$) on the system performance.
We assume that the BS is in the near-field of the RIS, and the RIS has $24 \times 24$ reflecting elements. We can observe from
this figure that the larger the size of the unit cells of the RIS, the better the system performance.
This is due to the fact that the total size of the RIS increases
when the size of the unit increases while keeping fixed the number of unit cells.

\begin{figure}[t]
	\begin{minipage}[t]{0.5\textwidth}
		\centering
		\includegraphics[scale=0.49]{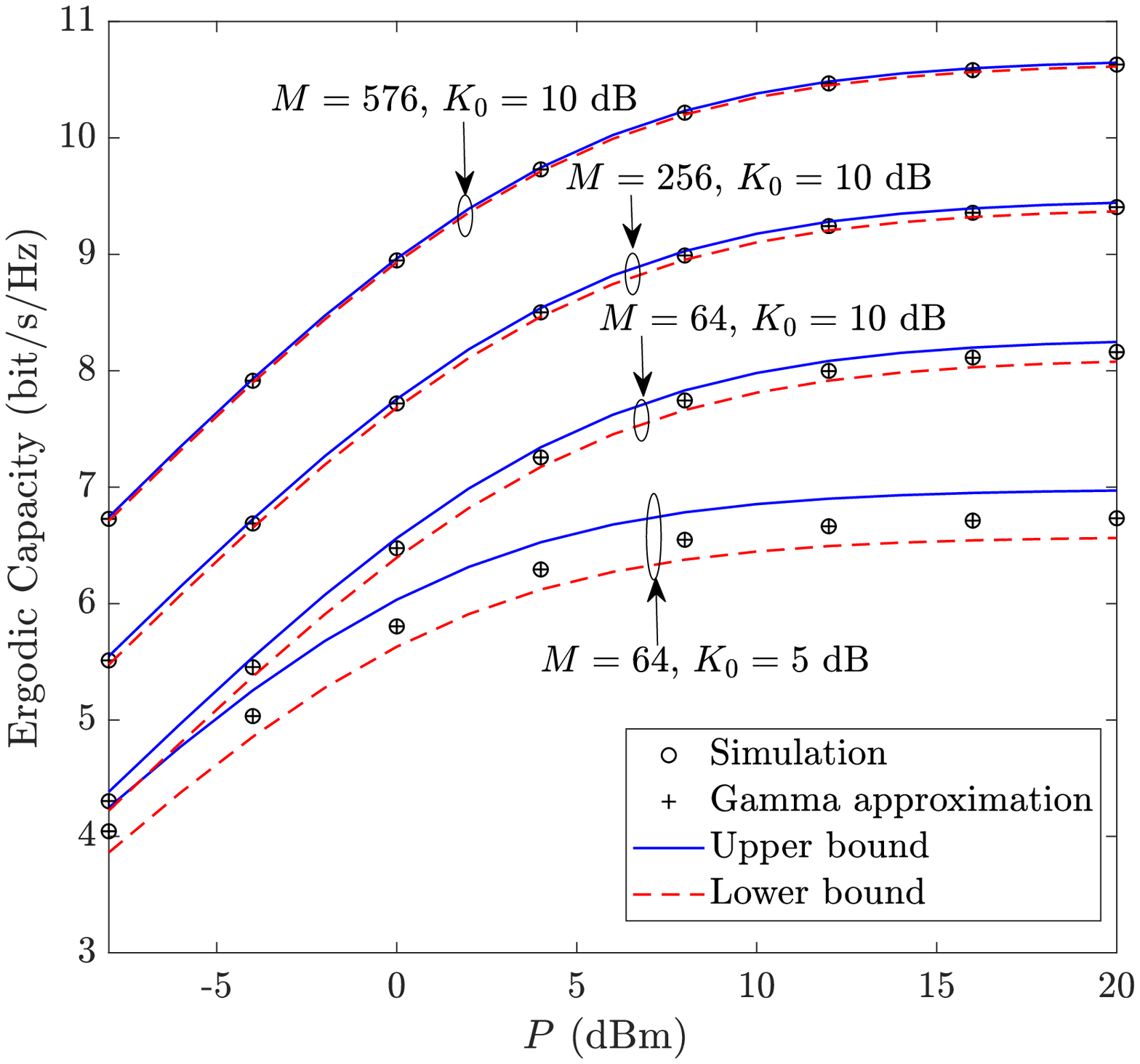}
		\caption{EC versus $P$ for different values of $K_0$ for centralized and distributed RIS deployments (${\rho _0} = 0.95$,
${\rho _\mathrm{C}} = 0.9$, ${\rho _{\mathrm{D}1}} = {\rho _{\mathrm{D}2}}=0.9$).\label{Figure_2}}
	\end{minipage}
	\qquad
	\begin{minipage}[t]{0.5\textwidth}
		\centering
          \includegraphics[scale=0.5]{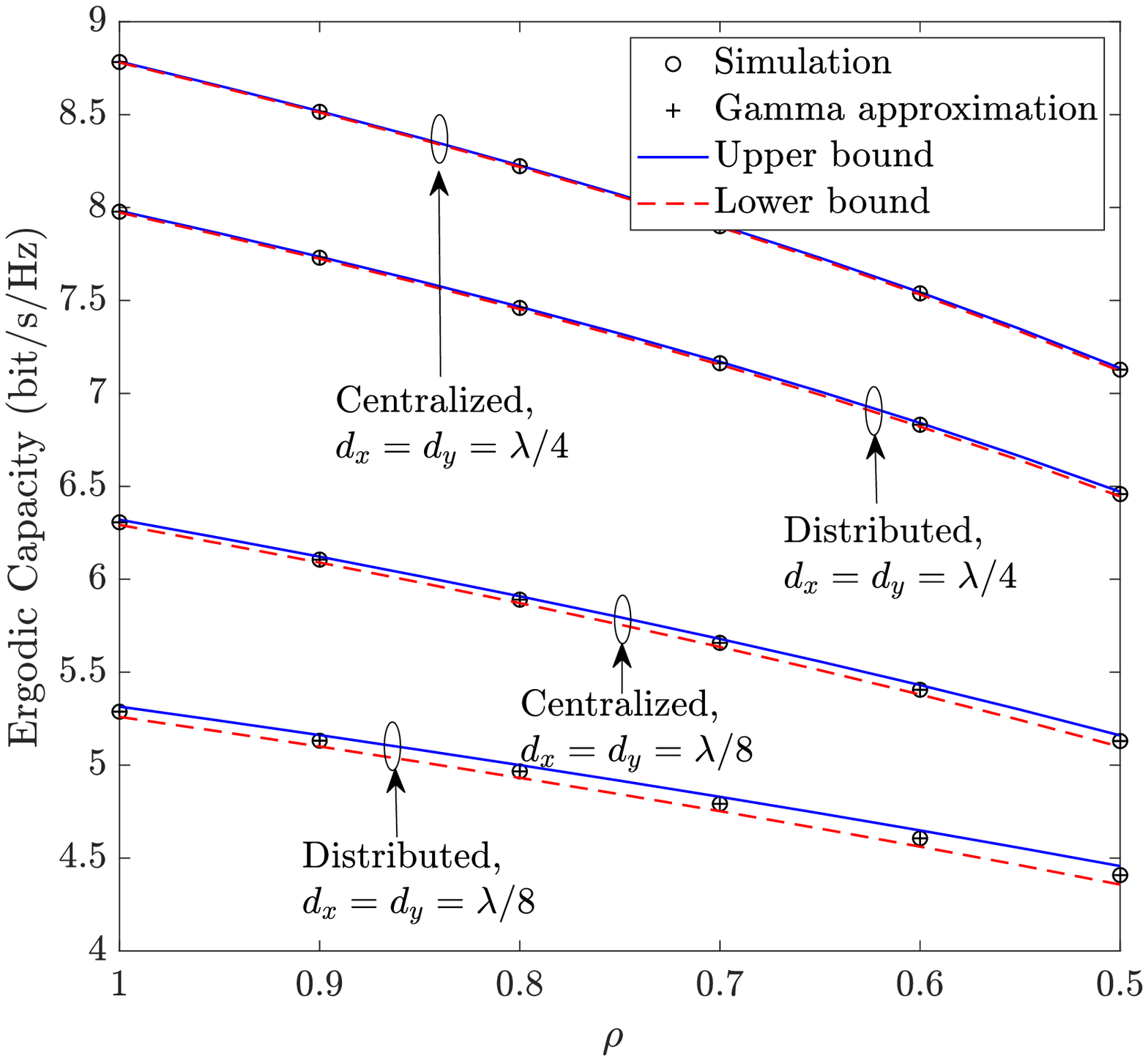}
		\caption{EC versus $\rho$ for different sizes of reflecting elements of RIS for centralized and distributed RIS deployments ($\rho_0=0.95$). \label{Figure_3}}
	\end{minipage}
\end{figure}

\begin{figure}[t]
	\begin{minipage}[t]{0.5\textwidth}
		\centering
		\includegraphics[scale=0.5]{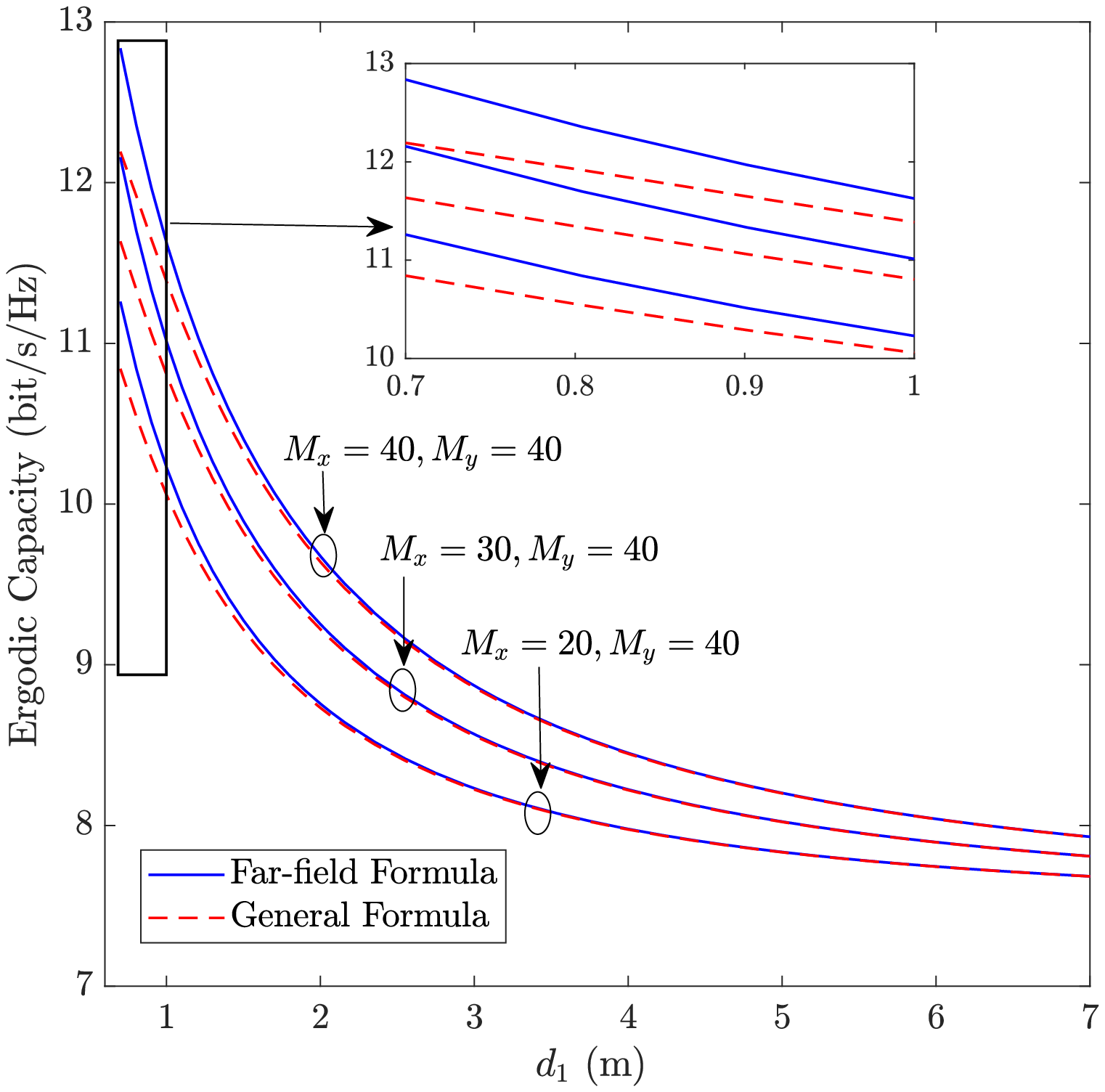}
		\caption{Comparison of the EC performance in the near- and far-field cases for the centralized deployment
(${d_x} = {d_y} = \frac{\lambda }{8}$, ${\rho _0} = 0.95$, ${\rho _\mathrm{C}} = 0.9$).\label{Figure_4}}
	\end{minipage}
	\qquad
	\begin{minipage}[t]{0.5\textwidth}
		\centering
		\includegraphics[scale=0.5]{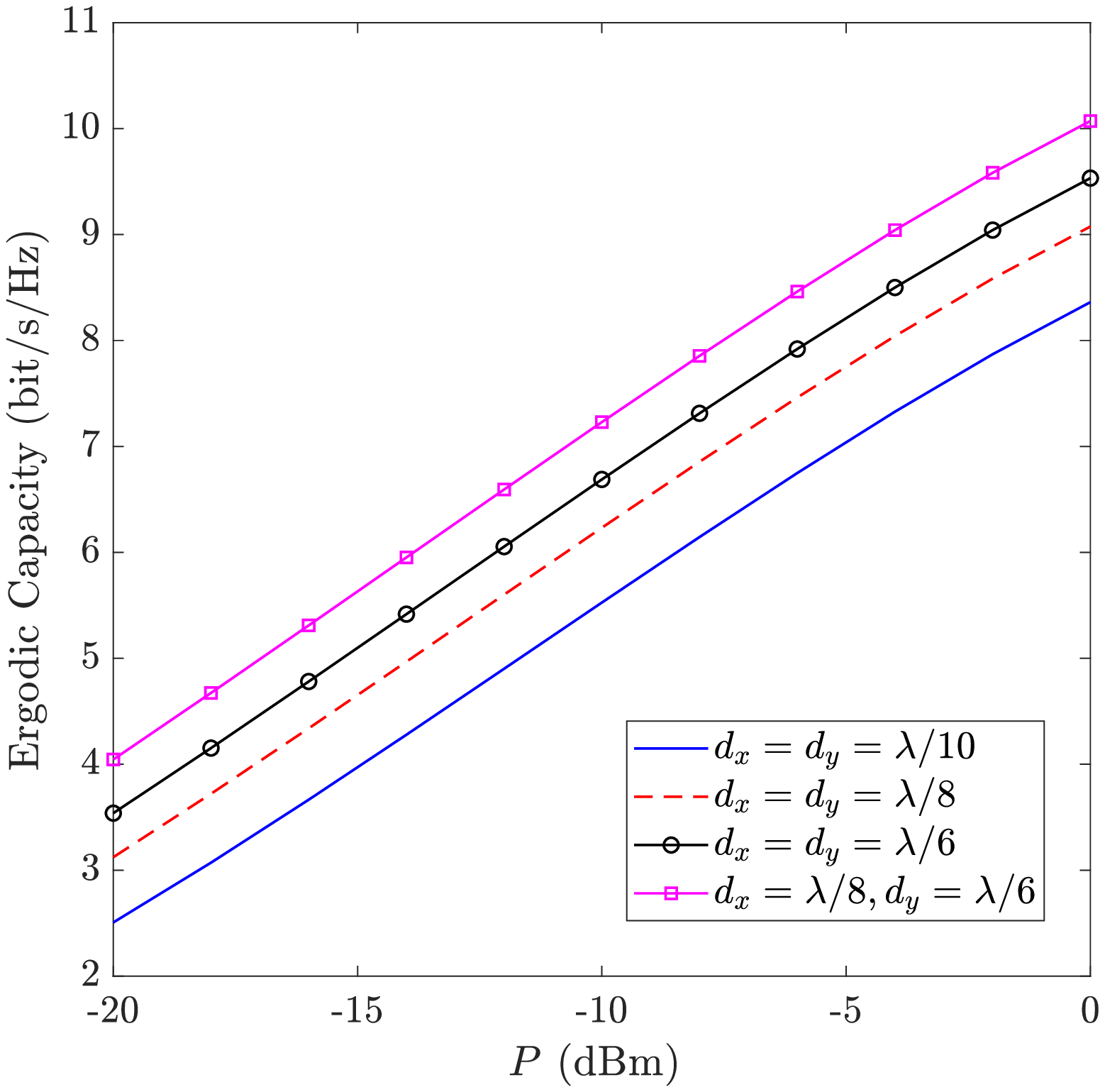}
		\caption{EC versus $P$ for a different size of the unit cells for the centralized RIS deployment in the near-field case
(${\rho _0} = 0.95$, ${\rho _\mathrm{C}} = 0.9$, ${M_x} = {M_y} = 24$). \label{Figure_5}}
	\end{minipage}
\end{figure}

\begin{figure}[t]
	\begin{minipage}[t]{0.5\textwidth}
		\centering
		\includegraphics[scale=0.5]{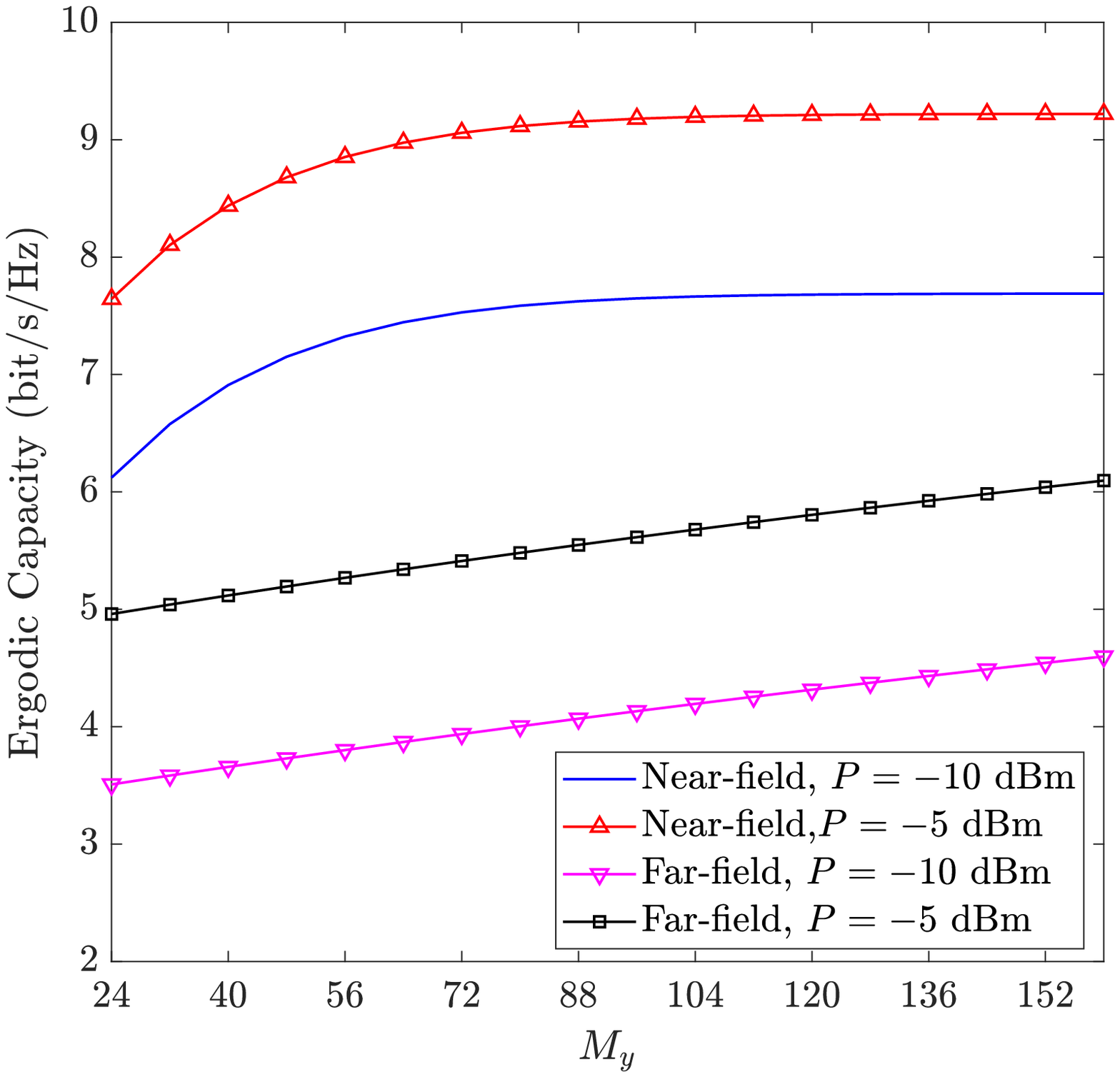}
		\caption{EC versus $M_y$ for different values of the transmit power in the near- and far-field regions for the centralized RIS deployment
(${M_x} = 24$, ${\rho _0} = 0.95$, ${\rho _\mathrm{C}} = 0.9$).\label{Figure_6}}
	\end{minipage}
	\qquad
	\begin{minipage}[t]{0.5\textwidth}
		\centering
		\includegraphics[scale=0.5]{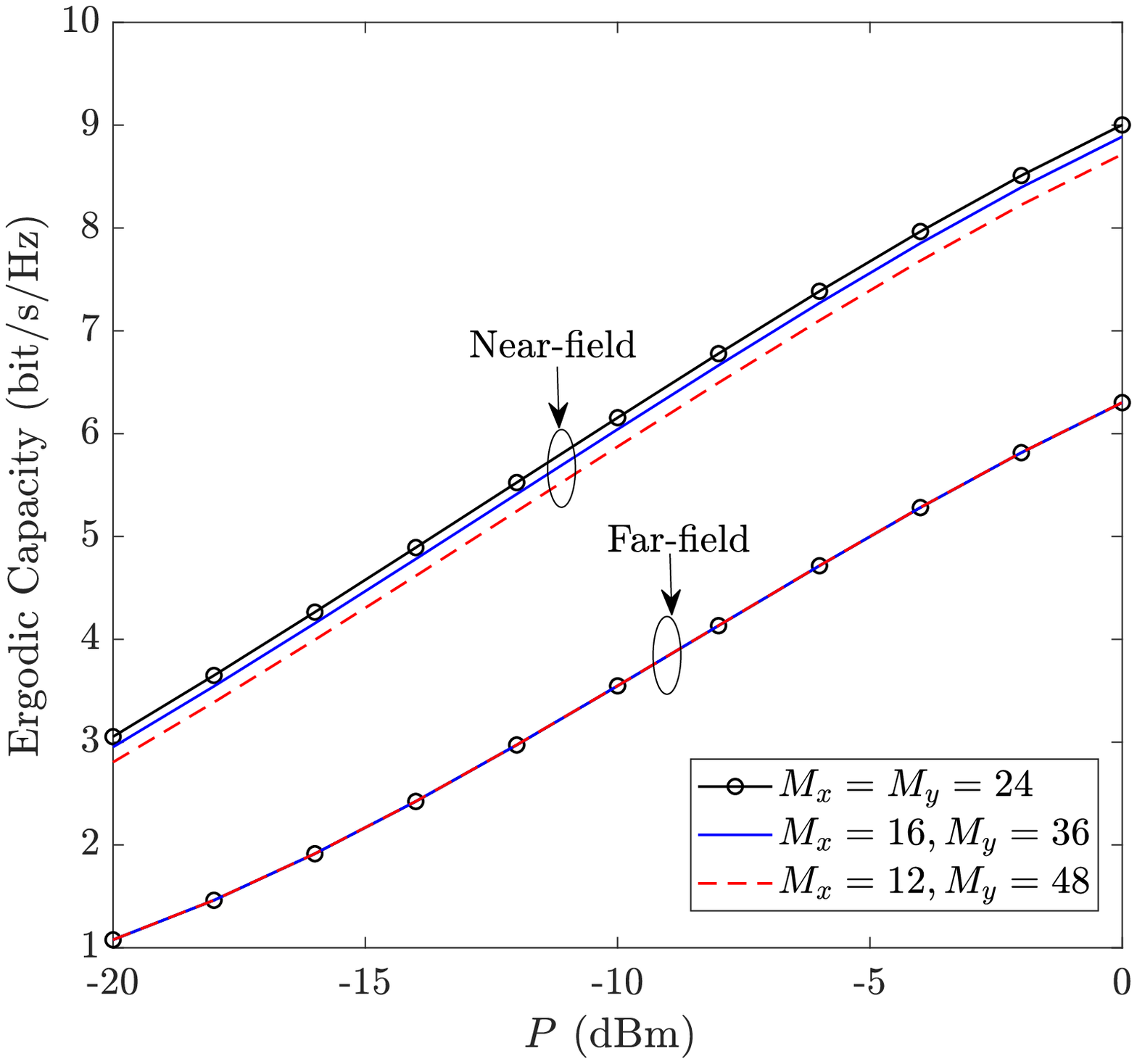}
		\caption{Comparison of the EC performance for different shapes of the RIS in the near- and
far-field regimes (${\rho _0} = 0.95$, ${\rho _\mathrm{C}} = 0.9$). \label{Figure_7}}
	\end{minipage}
\end{figure}

%
%
\begin{figure}[htbp]
\centering
\subfigure[Top view of the two RIS deployments]
{
	\begin{minipage}{9cm}
	\centering
	\includegraphics[scale=0.5]{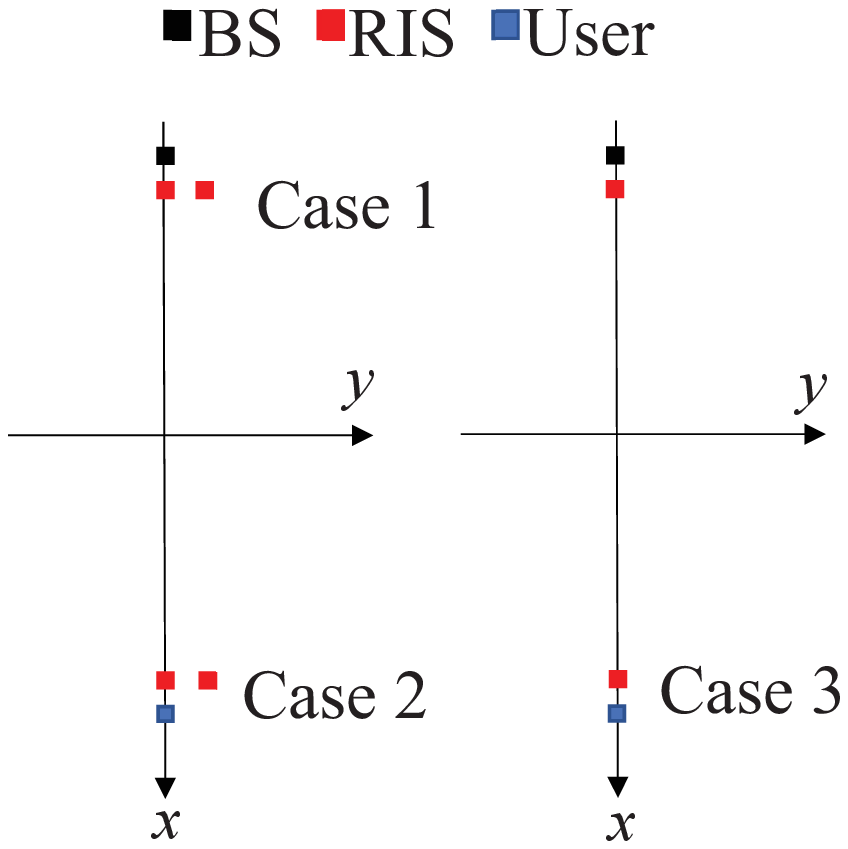}
	\end{minipage}
}
\subfigure[EC of centralized and distributed RIS-aided systems ($\rho_\mathrm{C}=0.9$). ]
{
	\begin{minipage}{9cm}
	\centering
	\includegraphics[scale=0.5]{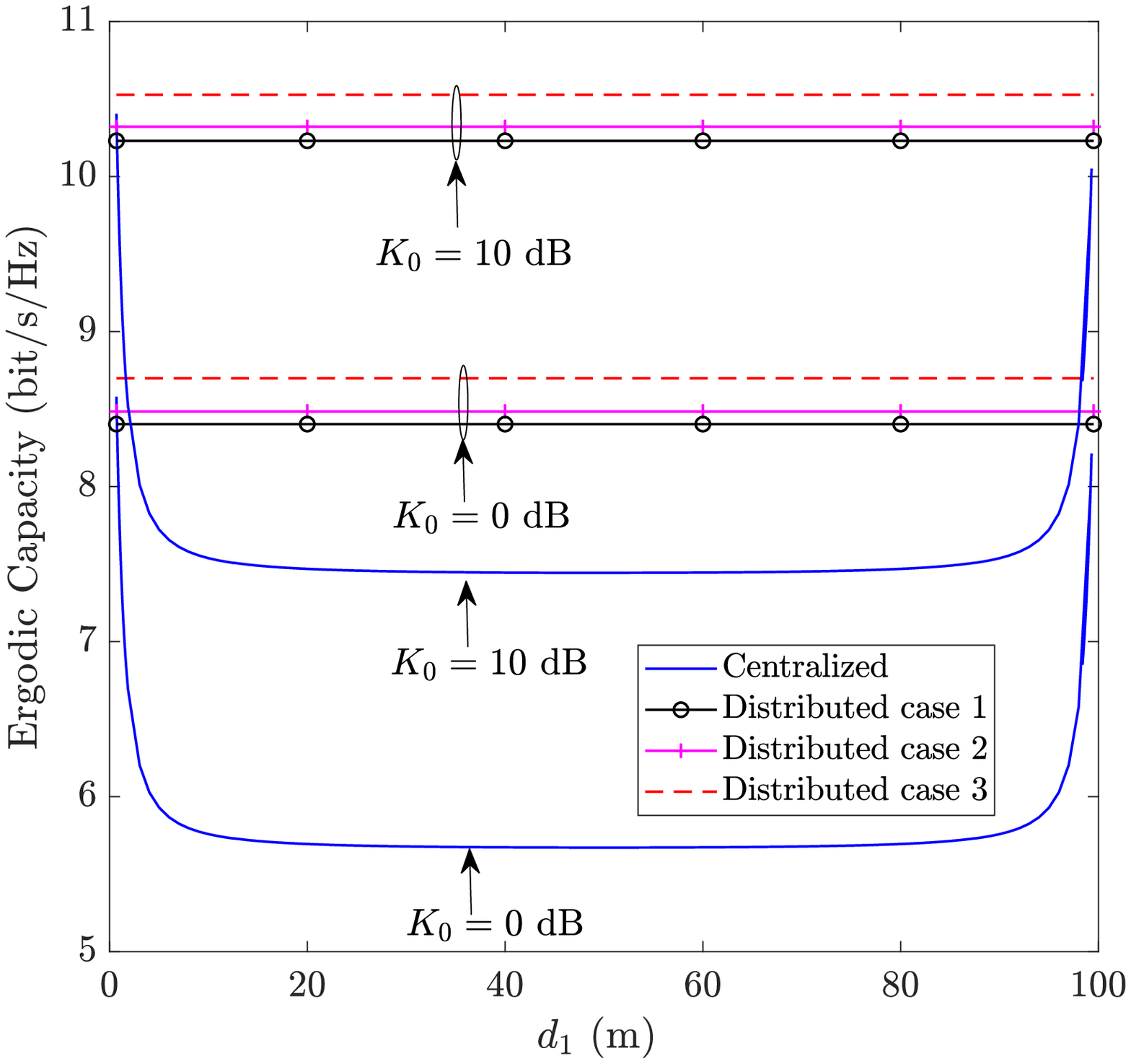}
	\end{minipage}
}
\subfigure[EC of centralized and distributed RIS-aided systems ($\rho_\mathrm{C}=1$). ]
{

\begin{minipage}{9cm}
	\centering
	\includegraphics[scale=0.5]{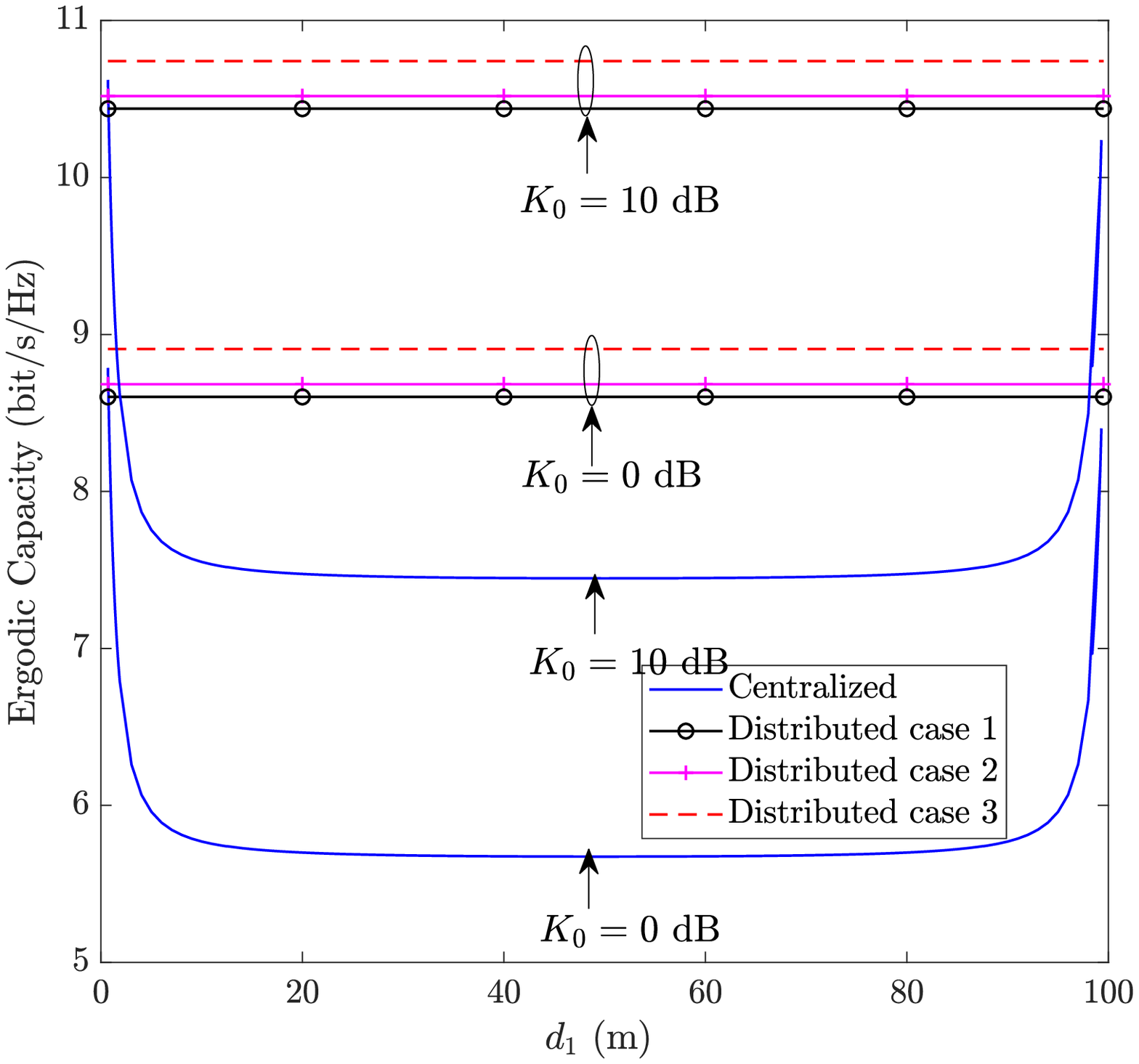}
	\end{minipage}
}
\caption{Comparison of the EC performance for different deployment strategies.}
\end{figure}

Figure 6 shows the EC in the near- and far-field regimes for the centralized RIS deployment
as a function of the total number of RIS reflecting elements. We set ${M_x}=24 $ fixed and increase $M_y$ linearly.
In the near-field regime, the RIS is located in $\left( { - 49.5,0,9.5} \right)$, and in the far-field regime,
the location of the RIS is $\left( { 0,0,9.5} \right)$.
We observe that when the BS is in the far-field of the RIS,
the EC increases with an increase of the number of RIS reflecting elements.
When the BS is in the near-field of the RIS, on the other hand,
the EC first increases with the number of RIS reflecting elements and then tends towards a  constant limit.
This performance trend can be explained as follows: If the BS and the receiver are steered towards the center of the RIS,
the RIS reflecting elements that are closer to the edge of the RIS
experience a more severe path loss compared to those that are closer to the
center of the RIS. Therefore, their contribution to the EC is not significant \cite{danufane2021path}, \cite{tang2021path}.

In Fig. 7, we analyze how different shapes of the RIS affect the EC with a fixed total number of reflecting elements of the RIS.
As can be readily observed, the more concentrated the reflecting elements installed on the RIS,
i.e., the closer the shape of the RIS to a square is,
the better the system performance in the near-field regime.
For example, if $P\!=\!-10$ dBm, setting $M_x \!=\!M_y=24$ results in $1.87\% $ and $4.78\% $ of
improvement of the EC compared to $M_x=16, M_y=36$ and $M_x=12, M_y=48$, respectively.
In addition, we observe that the shape of the RIS has no impact on the EC in the far-field regime,
because all the elements experience approximately the same path loss in the far-field regime,
while in the near-field case the path loss of each element is different.

In Fig. 8, we analyze the impact on the EC of the centralized and distributed deployments.
As for the centralized deployment, the single RIS is moved along the $x$-axis from the BS to the user.
As for the distributed deployment, we consider the following three different cases, as illustrated in Fig. 8 (a).
In case 1 and case 2, the two RISs are located near the BS and the user, respectively,
and the distance between the two RISs is $0.5$ m.
In case 3, on the other hand, one RIS is located near the BS and the other RIS is located near the user.
We observe from Fig. 8 (b) and Fig. 8 (c) that the EC degrades as $\rho$ decreases,
since a smaller value of $\rho$ implies a less accurate CSI.
Furthermore, in all distributed deployment cases, the EC is the largest in case 3. On the other hand, the scenario where the RIS is located near the user yields a better EC than the case where
the RIS is located near the BS.
In addition, it is found that the distributed deployment outperforms the centralized deployment.
The two deployments provide almost the same performance when a single RIS is located either near the BS or the user.
Moreover, the EC decreases as $K_0$ decreasing, and the conclusions of the comparison between the centralized and distributed deployments remain the same regardless of the values $K_0$.
\section{Conclusion}
In this paper, we have studied the ergodic capacity of centralized and distributed RIS-aided communication systems.
Considering the effects of near-field/far-field propagation conditions and outdated CSI,
we derived accurate closed-form approximations for the ergodic capacity.
Moreover, tight lower and upper bounds for the ergodic capacity were derived.
Our analysis reveals that the system performance improves with the transmit power,
the Rician-$K$ factor, the outdated CSI coefficient and the size of the reflecting elements.
Furthermore, the numerical results show that a distributed RIS-aided system usually outperforms a centralized RIS-aided system,
and that they provide almost the same ergodic capacity if a single RIS is deployed near the transmitter or near the receiver.

\setcounter{equation}{0}
\renewcommand{\theequation}{\thesection.\arabic{equation}}
\begin{appendices}
\section{Proof of Corollary \ref{coro:Gamma} }
Denote ${U_m} \!\!\buildrel \Delta \over = \!\!\sqrt {\!\beta _m^{ - 1}}
\left| {{{{\rm{\hat g}}}_{{\rm{C}}m}}} \right|\!\left| {{{\rm{h}}_{{\rm{C}}m}}} \right|$,
$T\! \buildrel \Delta \over = {\rho _{\rm{C}}}\!\!\sum\limits_{m = 1}^M \!\!{{U_m}} $,
$Z{\rm{ }} \buildrel \Delta \over = T + {\rho _0}\left| {{{{\rm{\hat g}}}_0}} \right|$,
$R \buildrel \Delta \over = {Z^2}$,
we have
 \begin{align}
 \mathbb{E}\left( {{U_m}} \right) = \sqrt {\beta _m^{ - 1}} {\Omega _1}{\Omega _2},\nonumber
 \end{align}
 \begin{equation}
\!\! \mathbb{E}\left( {U_m^2} \right) = \beta _m^{ - 1}.\nonumber
 \end{equation}
 Similarly, we can obtain the moments of $T$, $Z$ as
\begin{equation}
\mathbb{E}\left( T \right) = \mathbb{E}\left( {{\rho _{\rm{C}}}\sum\limits_{m = 1}^M {{U_m}} } \right)
= {\rho _{\rm{C}}}\sum\limits_{m = 1}^M {\sqrt {\beta _m^{ - 1}} {\Omega _1}{\Omega _2}}  \nonumber,
\end{equation}
 \begin{align}
&\mathbb{E}\left( {{T^2}} \right) = \mathbb{E}\left( {{{\left( {{\rho _{\rm{C}}}
\sum\limits_{m = 1}^M {{U_m}} } \right)}^2}} \right)\nonumber\\
&= \mathbb{E}\left( {\rho _{\rm{C}}^2\sum\limits_{m = 1}^M {U_m^2 + \rho _{\rm{C}}^2\sum\limits_{m = 1}^M
 {\sum\limits_{k = 1,k \ne m}^M {{U_m}{U_k}} } } } \right)\nonumber\\
& = \rho _{\rm{C}}^2\sum\limits_{m = 1}^M {\beta _m^{ - 1} + } \rho _{\rm{C}}^2\sum\limits_{m = 1}^M
 {\sum\limits_{k = 1,k \ne m}^M {\sqrt {\beta _m^{ - 1}\beta _k^{ - 1}} {{\left( {{\Omega _1}{\Omega _2}} \right)}^2}} }\nonumber,
  \end{align}
\begin{align}
\mathbb{E}\!\left( Z \right) \!\!=\!\! \mathbb{E}\!\left( {T \!+\!\! {\rho _0}\!\left| {{{{\rm{\hat g}}}_0}} \right|} \right) \!\!
 =\!\! {\rho _{\rm{C}}}\!\!\!\sum\limits_{m = 1}^M\!\! {\sqrt {\!\beta _m^{ - 1}} {\Omega _1}{\Omega _2}}
  \!+\!\! \sqrt {\!\beta _0^{ - 1}} {\rho _0}{\Omega _0}\nonumber,
\end{align}
\begin{align}
\label{EZ}
&\mathbb{E}\left( {{Z^2}} \right) = \mathbb{E}\left( {{{\left( {T + {\rho _0}\left| {{{{\rm{\hat g}}}_0}} \right|} \right)}^2}} \right) = \left( {{T^2} + \rho _0^2{{\left| {{{{\rm{\hat g}}}_0}} \right|}^2} + 2T{\rho _0}\left| {{{{\rm{\hat g}}}_0}} \right|} \right)\nonumber\\
& = \rho _{\rm{C}}^2\!\!\sum\limits_{m = 1}^M \!{\beta _m^{ - 1} + } \rho _{\rm{C}}^{\rm{2}}\!\sum\limits_{m = 1}^M \!{\sum\limits_{k = 1,k \ne m}^M \!\!{\sqrt {\beta _m^{ - 1}\beta _k^{ - 1}} {{\left( {{\Omega _1}{\Omega _2}} \right)}^2}} } \!\! +\!\! \beta _0^{ - 1}\rho _0^2 \!\!\nonumber\\
&+\!\! 2\sqrt {\beta _0^{ - 1}} {\rho _{\rm{C}}}{\rho _0}\sum\limits_{m = 1}^M {\sqrt {\beta _m^{ - 1}} {\Omega _0}{\Omega _1}{\Omega _2}}.
\end{align}
Then, with the aid of \cite[Sec. 2.2.2]{primak2005stochastic}, the PDF of $Z$ can be approximated with a Gamma distribution.
Therefore, \eqref{PDFofZ_cen} is obtained and the proof is completed.

\section{Proof of Corollary \ref{coro: EC_cen_general} }
\setcounter{equation}{0}
\renewcommand{\theequation}{\thesection.\arabic{equation}}
We first consider the derivation of ${\sigma _{{\rm{Ceff}}}^2}$. Using \eqref{received_signal_eff}, we obtain
\begin{align}
\label{A1}
&\sigma _{{\rm{Ceff}}}^2{\rm{ }} \buildrel \Delta \over =\mathbb{E }\left( {{{\left| {{n_{{\rm{Ceff}}}}} \right|}^2}} \right) = P\bar \rho _{\rm{C}}^2\mathbb{E}\left( {{{\left| {{{\bf{\omega }}_{\rm{C}}}{\bf{B}}{{\bf{\Phi }}_{\rm{C}}}{{\bf{h}}_{\rm{C}}}} \right|}^2}} \right) \nonumber\\
&+ P\bar \rho _{\rm{0}}^2\beta _0^{ - 1}\mathbb{E}\left( {{{\left| {{\omega _0}} \right|}^2}} \right) + \mathbb{E}\left( {{{\left| {{n_0}} \right|}^2}} \right).
\end{align}
In particular, we have
\begin{equation}
\label{A2}
\mathbb{E}\left( {{{\left| {{n_0}} \right|}^2}} \right) = \sigma _0^2,
\end{equation}
\begin{equation}
\label{A3}
\mathbb{E}\left( {{{\left| {{\omega _0}} \right|}^2}} \right) = \sigma _{{{{\rm{\hat h}}}_{\rm{0}}}}^2 = 1 - \Omega _0^2,
\end{equation}

Furthermore, we have
\begin{align}
\label{A4}
&\mathbb{E}\left( {{{\left| {{\bm{\omega} _{\rm{C}}}\bm{{\rm B}}{\bm{\Phi} _{\rm{C}}}{\mathbf{h}_{\rm{C}}}} \right|}^2}} \right)
= \mathbb{E}\left( {{{\left| {\sum\limits_{m = 1}^M {\frac{{{\omega _{{\rm{C}}m}}{\mathrm{h}_{{\rm{C}}m}}{e^{j{\varphi _{{\rm{C}}m}}}}}}
{{\sqrt {{\beta _m}} }}} } \right|}^2}} \right)\nonumber\\
& = \mathbb{E}\left( {\left| {\sum\limits_{m = 1}^M {\sum\limits_{k = 1}^M {\frac{{{\omega _{{\rm{C}}m}}{\mathrm{h}_{{\rm{C}}m}}
{e^{j{\varphi _{{\rm{C}}m}}}}}}{{\sqrt {{\beta _m}} }}\frac{{\omega _{{\rm{C}}k}^*\mathrm{h}_{{\rm{C}}k}^*{e^{ - j{\varphi _{{\rm{C}}k}}}}}}
{{\sqrt {{\beta _k}} }}} } } \right|} \right)\nonumber\\
& = \underbrace {\sum\limits_{m = 1}^M {\frac{{\mathbb{E}\!\!\left({{{\left| {{\omega _{{\rm{C}}m}}} \right|}^2}} \!\right)
\mathbb{E}\!\left( \!{{{\left| {{\mathrm{h}_{{\rm{C}}m}}} \right|}^2}}\right)}}{{{\beta _m}}}} }_{{I_1}}\nonumber\\
&+\!\! \underbrace {\sum\limits_{m = 1}^M \!{\sum\limits_{k = 1,k \ne m}^M\!\!\!\!\! {\mathbb{E}\!\left(\!\!
{\frac{{{\omega _{{\rm{C}}m}}{\mathrm{h}_{{\rm{C}}m}}\omega _{{\rm{C}}k}^*\mathrm{h}_{{\rm{C}}k}^*}}
{{\sqrt {{\beta _m}{\beta _k}} }}} \!\right){e^{j\left( {{\varphi _{{\rm{C}}m}} - {\varphi _{{\rm{C}}k}}} \right)}}} } }_{{I_2}}.
\end{align}
Since ${{\bm{\omega} _{\rm{C}}}}$, and ${{\mathbf{h}_{\rm{C}}}}$ are independent of each other and
$\mathbb{E}\left( {{\omega _{{\rm{C}}m}}} \right) = 0$, we obtain ${I_2} = 0$ and
\begin{align}
\label{A5}
\!\!\!\!\!\!\!{I_1} \!=\!\!\! \sum\limits_{m = 1}^M \!\!{\sigma _{{\mathrm{{\hat g}}_{{\rm{C}}m}}}^2\!\!\!
\left(\! {\sigma _{{\mathrm{h}_{{\rm{C}}m}}}^2 \!\!+\!\! {\left( {\mathbb{E}{{\left( {\left| {{{\rm{h}}_{{\rm{C}}m}}} \right|} \right)}^2}} \right)}}
\!\! \right)\!\beta _m^{ - 1}}   \!=\! \left( \!{1 \!\!-\! \Omega _2^2} \right)\!\!\!\sum\limits_{m = 1}^M \!{\beta _m^{ - 1}},
\end{align}

\noindent By substituting \eqref{A5} and ${I_2} = 0$ into \eqref{A4}, we arrive at
\begin{equation}
\label{A6}
\mathbb{E}\left( {{{\left| {{\bm{\omega} _{\rm{C}}}\bm{{\rm B}}{\bm{\Phi} _{\rm{C}}}{\mathbf{h}_{\rm{C}}}} \right|}^2}} \right)
= \left( {1 - \Omega _2^2} \right)\sum\limits_{m = 1}^M {\beta _m^{ - 1}}.
\end{equation}
By substituting \eqref{A2}, \eqref{A3} and \eqref{A6} into \eqref{A1}, we prove \eqref{sigma_Ceff}.

\noindent Now, we turn our attention to the derivation of the EC in \eqref{EC_cen_general}.
From \eqref{EC_Def} and employing the integration by parts method, the EC can be rewritten as
\begin{equation}
\label{C_cen_CDF}
{{\bar C}} = \frac{1}{{\ln 2}}\int_0^\infty  {\frac{{1 - {F_{{\gamma}}}
\left( \gamma  \right)}}{{1 + \gamma}}} d\gamma.
\end{equation}

\noindent Substituting \eqref{CDF_gamma_Cmax} into \eqref{C_cen_CDF} and using
\cite[Eq. (07.34.03.0271.01)]{Wolfram}, \cite[Eq. (07.34.03.0613.01)]{Wolfram}
together with \cite[Eq. (07.34.21.0013.01)]{Wolfram}, the EC can be obtained as in \eqref{EC_cen_general}.

\section{Proof of Corollary \ref{coro:Cen_EC_bound} }
\setcounter{equation}{0}
\renewcommand{\theequation}{\thesection.\arabic{equation}}
By using \eqref{SNR_cen_max}, we obtain
\begin{align}
\mathbb{E}\left( {{\gamma _{{\rm{Cmax}}}}} \right) = {\gamma _{{\rm{Cteff}}}}\mathbb{E}\left( {{Z^2}} \right).
\end{align}
\noindent Employing \eqref{EZ}, we obtain the upper bound for the EC in \eqref{EC_cen_ub}.
According to \eqref{PDFofZ_cen} and utilizing ${\gamma _{{\rm{Cmax}}}} = {\gamma _{{\rm{Cteff}}}}{Z^2}$,
we obtain the PDF of ${{\gamma _{{\rm{Cmax}}}}}$ as
\begin{align}
\!\!\!{f_{{\gamma _{{\rm{Cmax}}}}}}\!\left( \gamma  \right) \!=\! \frac{{b_{Z1}^{{a_{Z1}}}}}{{2\gamma _{{\rm{Cteff}}}^{{{{a_{Z1}}}
\mathord{\left/
 {\vphantom {{{a_{Z1}}} 2}} \right.
 \kern-\nulldelimiterspace} 2}}\Gamma \!\left( {{a_{Z1}}} \right)}}{\left( {\sqrt \gamma  } \right)^{{a_{Z1}} - 2}}
 {e^{ - {b_{Z1}}\sqrt {\frac{\gamma }{{{\gamma _{{\rm{Cteff}}}}}}} }}.
\end{align}
With the aid of $\mathbb{E}\left( {{\gamma _{{\rm{Cmax}}}}} \right) = \int_0^\infty  {\gamma {f_{{\gamma _{{\rm{Cmax}}}}}}
\left( \gamma  \right)} d\gamma $, $\mathbb{E}\left( {\gamma _{{\rm{C max}} }^2} \right) = \int_0^\infty
{{\gamma ^2}{f_{{\gamma _{{\rm{Cmax}}}}}}\left( \gamma  \right)} d\gamma $ and \cite[3.326.3]{Table},
we derive the mean and variance of ${{\gamma _{{\rm{Cmax}}}}}$ as
\begin{align}
\mathbb{E}\left( {{\gamma _{{\rm{Cmax}}}}} \right) \!&=\! \frac{{b_{Z1}^{{a_{Z1}}}}}{{2\gamma _{{\rm{Cteff}}}^{{{{a_{Z1}}}
\mathord{\left/
 {\vphantom {{{a_{Z1}}} 2}} \right.
 \kern-\nulldelimiterspace} 2}}\Gamma\! \left(\! {{a_{Z1}}}\! \right)}}\!\int_0^\infty \!\!\!\!
 {\gamma {{\left( {\sqrt \gamma  } \right)}
 ^{{a_{Z1}} \!- \!2}}{e^{ - {b_{Z1}}\sqrt {\frac{\gamma }{{{\gamma _{{\rm{Cteff}}}}}}} }}} d\gamma \nonumber\\
& = {\gamma _{{\rm{Cteff}}}}\frac{{\Gamma \left( {{a_{Z1}} + 2} \right)}}{{b_{Z1}^2\Gamma \left( {{a_{Z1}}} \right)}},
\end{align}
\begin{align}
\mathbb{E}\!\left( {\gamma _{{\rm{Cmax }}}^2} \right) \!&=\! \frac{{b_{Z1}^{{a_{Z1}}}}}{{2\gamma _{{\rm{Cteff}}}^{{{{a_{Z1}}}
\mathord{\left/
 {\vphantom {{{a_{Z1}}} 2}} \right.
 \kern-\nulldelimiterspace} 2}}\Gamma\! \left(\! {{a_{Z1}}} \!\right)}}\!\int_0^\infty \!\!\!\! {{\gamma ^2}{{\left( \!{\sqrt \gamma  }
  \right)}^{{a_{Z1}}\! -\! 2}}{e^{ - {b_{Z1}}\sqrt {\frac{\gamma }{{{\gamma _{{\rm{Cteff}}}}}}} }}} d\gamma \nonumber\\
 &= \gamma _{{\rm{Cteff}}}^2\frac{{\Gamma \left( {{a_{Z1}} + 4} \right)}}{{b_{Z1}^4\Gamma \left( {{a_{Z1}}} \right)}},
\end{align}
Finally, by utilizing the relation ${\rm{Var}}\left( {{\gamma _{{\rm{Cmax}}}}} \right) =\mathbb{ E}\left( {\gamma _{{\rm{Cmax}} }^2}
 \right) - {\left( {{\rm{\mathbb{E}}}\left( {{\gamma _{{\rm{Cmax}}}}} \right)} \right)^2}$, we obtain the
 variance in \eqref{Variance}.

\section{Proof of Corollary \ref{coro:gamma_Dis} }
\setcounter{equation}{0}
\renewcommand{\theequation}{\thesection.\arabic{equation}}
By denoting ${X_{lm}} \buildrel \Delta \over = \sqrt {\beta _{lm}^{ - 1}} \left| {{\mathrm{{\hat g}}_{{\rm{D}}lm}}} \right|
\left| {{\mathrm{h}_{{\rm{D}}lm}}} \right|$,
 $Y \buildrel \Delta \over = \sum\limits_{l = 1}^L {\sum\limits_{m = 1}^{{M_l}} {{\rho _{{\rm{D}}l}}{X_{lm}}} } $
 and $H \buildrel \Delta \over = Y +  {\rho _0}\left| {{{{\rm{\hat g}}}_0}} \right|$, we have
\begin{align}
\mathbb{E}\left( {{X_{lm}}} \right) \!=\! \mathbb{E}\!\left(\!\! {\sqrt {\beta _{lm}^{ - 1}}
\left| {{\mathrm{{\hat g}}_{{\rm{D}}lm}}} \right|\left| {{\mathrm{h}_{{\rm{D}}lm}}} \right|} \!\right)
\!=\! \sqrt {\beta _{lm}^{ - 1}} {\Omega _{1l}}{\Omega _{2l}},
\end{align}
\begin{align}
\mathbb{E}\left( {X_{lm}^2} \right) = \beta _{lm}^{ - 1},
\end{align}
\begin{align}
&\mathbb{E}\left( Y \right) = \mathbb{E}\left( {\sum\limits_{l = 1}^L {\sum\limits_{m = 1}^{{M_l}}
{{\rho _{{\rm{D}}l}}{X_{lm}}} } } \right)\nonumber\\
& = \sum\limits_{l = 1}^L {\sum\limits_{m = 1}^{{M_l}}
{{\rho _{{\rm{D}}l}}\mathbb{E}\left( {{X_{lm}}} \right)} }
 = \sum\limits_{l = 1}^L {\sum\limits_{m = 1}^{{M_l}} {{\rho _{{\rm{D}}l}}\sqrt {\beta _{lm}^{ - 1}} {\Omega _{1l}}{\Omega _{2l}}} }.
\end{align}
\begin{align}
&\mathbb{E}\left( {{Y^2}} \right) = \mathbb{E}\left( {{{\left( {\sum\limits_{l = 1}^L
{\sum\limits_{m = 1}^{{M_l}} {{\rho _{{\rm{D}}l}}{X_{lm}}} } } \right)}^2}} \right)\nonumber\\
 &= \mathds{E}\left( {\sum\limits_{l = 1}^L {\sum\limits_{m = 1}^{{M_l}} {\rho _{{\rm{D}}l}^2X_{lm}^2} }
 + \sum\limits_{l = 1}^L {\sum\limits_{m = 1}^{{M_l}} {\sum\limits_{k = 1,k \ne m}^{{M_l}}
 {\rho _{{\rm{D}}l}^2{X_{lm}}{X_{lk}}} } } } \right.\nonumber\\
&\left. { + \sum\limits_{l = 1}^L {\sum\limits_{m = 1}^{{M_l}} {\left( {{\rho _{{\rm{D}}l}}
{X_{lm}}\sum\limits_{j = 1,j \ne l}^L {\sum\limits_{k = 1}^{{M_j}} {{\rho _{{\rm{D}}j}}{X_{jk}}} } } \right)} } } \right)\nonumber\\
& = \sum\limits_{l = 1}^L {\sum\limits_{m = 1}^{{M_l}} {\frac{{\rho _{{\rm{D}}l}^2}}{{{\beta _{lm}}}}} }
+ \sum\limits_{l = 1}^L {\sum\limits_{m = 1}^{{M_l}} {\sum\limits_{k = 1,k \ne m}^{{M_l}}
 {\frac{{\rho _{{\rm{D}}l}^2\Omega _{1l}^2\Omega _{2l}^2}}{{\sqrt {{\beta _{lm}}{\beta _{lk}}} }}} } } \nonumber\\
&+\sum\limits_{l = 1}^L {\sum\limits_{m = 1}^{{M_l}} {\left( {\frac{{{\rho _{{\rm{D}}l}}
{\Omega _{1l}}{\Omega _{2l}}}}{{\sqrt {{\beta _{lm}}} }}\sum\limits_{j = 1,j \ne l}^L
{\sum\limits_{k = 1}^{{M_j}} {\frac{{{\rho _{{\rm{D}}j}}{\Omega _{1j}}{\Omega _{2j}}}}{{\sqrt {{\beta _{lk}}} }}} } } \right)} }.
\end{align}

\begin{align}
\!\!\!\!&\mathbb{E}\!\left( H \right) \!=\! \mathbb{E}\left( {Y +
{\rho _0}\left| {{{{\rm{\hat g}}}_0}} \right|} \right)
=\!\! \sum\limits_{l = 1}^L \!{\sum\limits_{m = 1}^{{M_l}}\!\! {\sqrt {\!\beta _{lm}^{ - 1}}
{\rho _{{\rm{D}}l}}{\Omega _{1l}}{\Omega _{2l}}} } \! +\! \sqrt {\!\beta _0^{ - 1}} {\rho _0}{\Omega _0}.
\end{align}

\begin{align}
\label{CC9}
&\mathbb{E}\!\!\left(\! {{H^2}}\! \right) \!\!=\!\mathbb{ E}\!\left( {{{\left( {Y \!+\!
\! {\rho _0}\!\left| {{{{\rm{\hat g}}}_0}} \right|} \right)}^2}}\right) \!=\!\mathbb{ E}\!
\left( {{Y^2} \!+\! \beta _0^{ - 1}\!\rho _0^2 \!+\! 2Y \! {\rho _0}\!
\left| {{{{\rm{\hat g}}}_0}} \right|} \right)\nonumber\\
& = \sum\limits_{l = 1}^L {\sum\limits_{m = 1}^{{M_l}} {\frac{{\rho _{{\rm{D}}l}^2}}{{{\beta _{lm}}}}} }
+ \sum\limits_{l = 1}^L {\sum\limits_{m = 1}^{{M_l}} {\sum\limits_{k = 1,k \ne m}^{{M_l}}
{\frac{{\rho _{{\rm{D}}l}^2\Omega _{1l}^2\Omega _{2l}^2}}{{\sqrt {{\beta _{lm}}{\beta _{lk}}} }}} } }\nonumber \\
&+\sum\limits_{l = 1}^L {\sum\limits_{m = 1}^{{M_l}} {\left( {\frac{{{\rho _{{\rm{D}}l}}{\Omega _{1l}}
{\Omega _{2l}}}}{{\sqrt {{\beta _{lm}}} }}\sum\limits_{j = 1,j \ne l}^L {\sum\limits_{k = 1}^{{M_j}}
{\frac{{{\rho _{{\rm{D}}j}}{\Omega _{1j}}{\Omega _{2j}}}}{{\sqrt {{\beta _{jk}}} }}} } } \right)} }\nonumber \\
& + \frac{{2{\rho _0}}}{{\sqrt {{\beta _0}} }}\sum\limits_{l = 1}^L {\sum\limits_{m = 1}^{{M_l}}
{\frac{{{\rho _{{\rm{D}}l}}{\Omega _{1l}}{\Omega _{2l}}}}{{\sqrt {{\beta _{lm}}} }} + \frac{{\rho _0^2}}{{{\beta _0}}}} }.
\end{align}
Then, according to \cite[Sec. 2.2.2]{primak2005stochastic}, the PDF of $H$ can be tightly approximated by the Gamma
distribution which is characterized by two parameters
${a_{H1}} \buildrel \Delta \over = \frac{{{{\left( {\mathbb{E}\left( H \right)} \right)}^2}}}{{{\rm{Var}}\left( H \right)}}$,
${b_{H1}} \buildrel \Delta \over = \frac{{\mathbb{E}\left( H \right)}}{{{\rm{Var}}\left( H \right)}}$.
Then, \eqref{PDFofH} is proved.

\section{Proof of Corollary \ref{coro:EC_dis}}
\setcounter{equation}{0}
\renewcommand{\theequation}{\thesection.\arabic{equation}}
The variance of the equivalent noise ${{{{{n_{{\rm{Deff}}}}} }}}$ for the distributed deployment can be derived as
\begin{align}
\label{C1}
\sigma _{{\rm{Deff}}}^2{\rm{ }}& \buildrel \Delta \over =\mathbb{E} \left( {{{\left| {{n_{{\rm{Deff}}}}} \right|}^2}} \right) = P\sum\limits_{l = 1}^L  \bar \rho _{{\rm{D}}l}^2\mathbb{E}\left( {{{\left| {{{\bf{\omega }}_{{\rm{D}}l}}{{\bf{B}}_{{\rm{D}}l}}{{\bf{\Phi }}_{{\rm{D}}l}}{{\bf{h}}_{{\rm{D}}l}}} \right|}^2}} \right)\nonumber\\
&+ P\bar \rho _{\rm{0}}^2\beta _0^{ - 1}\mathbb{E}\left( {{{\left| {{\omega _0}} \right|}^2}} \right) + \mathbb{E}\left( {{{\left| {{n_0}} \right|}^2}} \right),
\end{align}
where $\mathbb{E}\left( {{{\left| {{n_0}} \right|}^2}} \right)$
and $\mathbb{E}\left( {{{\left| {{\omega _0}} \right|}^2}} \right)$ are given in \eqref{A2} and \eqref{A3}, respectively.
Moreover, we have
\begin{small}
\begin{align}
\label{C2}
&\mathbb{E}\left( {{{\left| {{\bm{\omega} _{{\rm{D}}l}}{\bm{{\rm B}}_{{\rm{D}}l}}
{\bm{\Phi} _{{\rm{D}}l}}{\mathbf{h}_{{\rm{D}}l}}} \right|}^2}} \right)= \mathbb{E}\!\!\left( {\sum\limits_{m = 1}^{{M_l}} \!\!{\frac{{{{\left| {{\omega _{{\rm{D}}lm}}{\mathrm{h}_{{\rm{D}}lm}}} \right|}^2}}}
{{\sqrt {{\beta _{lm}}} }}} } \right)\nonumber\\
 &+\mathbb{E}\left( {{{\left| {\sum\limits_{m = 1}^{{M_l}}
 {\sum\limits_{k \ne m}^{{M_l}} {\frac{{{\omega _{{\rm{D}}lm}}{\mathrm{h}_{{\rm{D}}lm}}
 \omega _{{\rm{D}}lk}^*\mathrm{h}_{{\rm{D}}lk}^*{e^{j\left( {{\varphi _{lm}} - {\varphi _{lk}}} \right)}}}}
 {{\sqrt {{\beta _{lm}}{\beta _{lk}}} }}} } } \right|}^2}} \right)\nonumber\\
& \!=\! \underbrace {\sum\limits_{m = 1}^{{M_l}} {\frac{{\mathbb{E}\!\!\left( {{{\left| {{\omega _{{\rm{D}}lm}}} \right|}^2}} \right)
\mathbb{E}\left( {{{\left| {{\mathrm{h}_{{\rm{D}}lm}}} \right|}^2}} \right)}}{{{\beta _{lm}}}}} }_{{I_3}}\nonumber\\
&+ \!\underbrace {\sum\limits_{m = 1}^{M - 1} {\sum\limits_{k = 1,k \ne m}^M {\mathbb{E}
\left( {\frac{{{\omega _{{\rm{D}}lm}}{\mathrm{h}_{\mathrm{D}lm}}
\omega _{{\rm{D}}lk}^*\mathrm{h}_{{\rm{D}}lk}^*}}{{\sqrt {{\beta _{lm}}{\beta _{lk}}} }}} \right)
{e^{j\left( {{\varphi _{lm}} - {\varphi _{lk}}} \right)}}} } }_{{I_4}}.
\end{align}
\end{small}
\noindent Due to the fact that ${{\bm{\omega} _{\mathrm{D}l}}}$, and ${{\mathbf{h}_{\mathrm{D}l}}}$ are independent of each other,
we obtain ${I_4}{\rm{ = }}0$ and
\begin{align}
\label{C3}
\!\!\!{I_3} \!\!=\!\!\! \sum\limits_{m = 1}^{{M_l}}\!\! {\beta _{lm}^{ - 1}\sigma _{{\mathrm{{\hat g}}_{{\rm{D}}lm}}}^2}\!\!
\left(\! {\sigma _{{\mathrm{h}_{{\rm{D}}lm}}}^2\! \!\!+\!\! \mathbb{E}{^2}\!
\left( {\left| {{\mathrm{h}_{{\rm{D}}lm}}} \right|} \right)}\! \right)\!\!
=\!\! \left(\! {1 \!-\! \Omega _{2l}^2} \right)\!\!\!\sum\limits_{m = 1}^{{M_l}}\!\! {\beta _{lm}^{ - 1}}.
\end{align}
Substituting \eqref{C2}, \eqref{C3}, ${I_4}{\rm{ = }}0$, \eqref{A2} and \eqref{A3} into \eqref{C1},
we arrive at \eqref{noise_variance_dis}.

\noindent Next, by exploiting a similar methodology as for the derivation of \eqref{CDF_gamma_Cmax}, we arrive at
\begin{equation}
\label{C10}
{F_{{\gamma _{{\rm{Dmax}}}}}}\left( \gamma  \right) = 1 - \frac{1}{{\Gamma \left( {{a_{H1}}} \right)}}
\Gamma \left( {{a_{H1}},{b_{H1}}\sqrt {\frac{\gamma }{{{\gamma _{{\rm{Dteff}}}}}}} } \right).
\end{equation}
Substituting \eqref{C10} into \eqref{C_cen_CDF} and with the aid of \cite[Eq. (07.34.03.0271.01)]{Wolfram},
\cite[Eq. (07.34.03.0613.01)]{Wolfram} and \cite[Eq. (07.34.21.0013.01)]{Wolfram},
we obtain \eqref{EC_dis_near_gamma}.

\section{Proof of Corollary \ref{coro:Dis_EC_bound}}
\setcounter{equation}{0}
\renewcommand{\theequation}{\thesection.\arabic{equation}}
We can rewrite \eqref{SNR_max_dis} as
\begin{align}
\mathbb{E}\left( {{\gamma _{{\rm{Dmax}}}}} \right) = {\gamma _{{\rm{Dteff}}}}\mathbb{E}\left( {{H^2}} \right).
\end{align}
Next, by employing \eqref{C9}, we obtain \eqref{EC_dis_ub}.
Then, $\mathrm{Var}\left( {{\gamma _{{\rm{Dmax}} }}} \right)$ is obtained by using the same
approach as for the derivation of \eqref{Variance}.

\end{appendices}


\bibliographystyle{IEEEtran}
\bibliography{IEEEabrv,Ref}

\end{document}